# The Next Layer: Augmenting Foundation Models with Structure-Preserving and Attention-Guided Learning for Local Patches to Global Context Awareness in Computational Pathology

Muhammad Waqas[1], Rukhmini Bandyopadhyay[1], Eman Showkatian[1], Amgad Muneer[1], Anas Zafar[1], Frank Rojas Alvarez[2,3], Maricel Corredor Marin[3], Wentao Li[1], David Jaffray[4], Cara Haymaker[3], John Heymach[5], Natalie I Vokes[6], Luisa Maren Solis Soto[3], Jianjun Zhang[5,6], Jia Wu[1,5] ✉

1. Department of Imaging Physics, MD Anderson Cancer Center, Houston, TX, USA
2. Anatomic Pathology, Northwestern University, Chicago, IL, USA
3. Translational Molecular Pathology, The University of Texas MD Anderson Cancer Center
4. Office of the Chief Technology and Digital Officer, The University of Texas MD Anderson Cancer Center, Houston, TX, USA
5. Department of Thoracic/Head and Neck Medical Oncology, MD Anderson Cancer Center, Houston, TX, USA
6. Department of Genomic Medicine, MD Anderson Cancer Center, Houston, TX, USA

**Corresponding Author**

Jia Wu, PhD

Department of Imaging Physics

Department of Thoracic/Head and Neck Medical Oncology

The University of Texas MD Anderson Cancer Center

1515 Holcombe Blvd

Houston, TX 77030, USA

Telephone: 713-563-2719

e-mail: jwu11@mdanderson.org

**Acknowledgements**

This work was supported by the generous philanthropic contributions to The University of Texas MD Anderson Lung Moon Shot Program, the MD Anderson Cancer Center Support Grant P30CA016672. This research was partially funded by the National Institutes of Health (NIH) grants R01CA276178 (N.I.V. and J.W.) and CPRIT RP240117 (J.W.). This work was also sponsored by generous philanthropic contributions from Mrs. Andrea Mugnaini and Dr. Edward L.C. Smith, as well as the Rexanna's Foundation for Fighting Lung Cancer, QIAC Partnership in Research (QPR) funding, and Permanent Health Funds.

**Author contributions**

M.W, R.B, and J.W conceptualized the clinical problem and overall study design. R.B, A.M, ES, W.L, acquired and processed the multi-center data. M.W, R.B, designed Eagle-Net. A.M, A.Z analyzed the EAGLE-Net model. M.W, R.B, A.M, A.Z, and J.W performed quality control of the data and algorithms. F.R.A and M.C.M performed data annotation. J.W, J.Z supervised the project. D.J, C.H, J.H, N.I.V, L.M.S.S contributed to the review and editing of the manuscript. All authors had access to the data presented in the manuscript and approved the final version for publication.

# Abstract

Foundation models have recently emerged as powerful feature extractors in computational pathology, yet they typically omit mechanisms for leveraging the global spatial structure of tissues and the local contextual relationships among diagnostically relevant regions—key elements for understanding the tumor microenvironment. Multiple instance learning (MIL) remains an essential next step following foundation model, designing a framework to aggregate patch-level features into slide-level predictions. We present EAGLE-Net, a structure-preserving, attention-guided MIL architecture designed to augment prediction and interpretability. EAGLE-Net integrates multi-scale absolute spatial encoding to capture global tissue architecture, a top-K neighborhood-aware loss to focus attention on local microenvironments, and background suppression loss to minimize false positives. We benchmarked EAGLE-Net on large pan-cancer datasets, including three cancer types for classification (10,260 slides) and seven cancer types for survival prediction (4,172 slides), using three distinct histology foundation backbones (REMEDIES, Uni-V1, Uni2-h). Across tasks, EAGLE-Net achieved up to 3% higher classification accuracy and the top concordance indices in 6 of 7 cancer types, producing smooth, biologically coherent attention maps that aligned with expert annotations and highlighted invasive fronts, necrosis, and immune infiltration. These results position EAGLE-Net as a generalizable, interpretable framework that complements foundation models, enabling improved biomarker discovery, prognostic modeling, and clinical decision support.

**Keywords**: Computational Pathology, Foundation Model, Multiple Instance Learning, Explainable AI.

# INTRODUCTION

The tumor is a tissue mass of abnormal cells that indicates the presence of cancer. It is a complex and evolving ecosystem shaped by selective pressures from its microenvironment, including immunological, metabolic, trophic, and therapeutic factors [1]. These forces influence the distribution, abundance, and functional orientation of different cellular components within the tumor microenvironment (TME), leading to phenotypic and spatial diversity known as intra-tumoral heterogeneity (ITH) [2]. ITH fosters the emergence of cancer cells that evade immune surveillance, undergo genetic evolution, and develop resistance to therapies [3]. Within the TME, specialized cellular "niches"—comprising diverse cell populations such as cancer, vasculature, immune, adipocytes, fibroblasts, nerve cells, and extracellular matrix components—create distinct habitats that drive tumor growth, invasion, metastasis, and influence treatment responses. The spatial arrangement of these niches—their proximity, boundaries, and cellular composition—encodes prognostic clues and vulnerabilities, and understanding the intricate interactions and spatial arrangement within these niches is essential for the development of more effective cancer therapies[4].

Beyond routine pathologist evaluation, computational pathology leverages machine learning tools on digitized hematoxylin and eosin (H&E) whole slide images (WSIs), enabling micrometer-resolution assessment across gigapixel-scale images [5,6]. Recent rising of foundation models trained through self-supervised learning on cropped patches provides powerful tools for feature extraction [7-9]. WSIs are typically represented as a set of patch-level embeddings, each patch treated as an instance and processed through the Multiple Instance Learning (MIL) framework. Under the hypothesis that only subset of instances are relevant for prediction, MIL learns to aggregate unannotated patch-level information to predict slide-level or patient-level outcomes [4,10-13].

While MIL models offer modest predictive performance and instance-level interpretability, they often neglect both global tissue-level architecture and the local spatial context of informative patches [11,14-18], leading to suboptimal prediction. Biologically speaking, tumor behavior is shaped not just by the presence of cellular niches but by their spatial organization and interaction patterns [19,20]. Standard positional encoding techniques, such as those proposed in vision transformers (ViT)[21-23], are intended for fixed-length sequences and fail to handle the variable number of patches in WSI analysis. Alternatives like pyramid-based encoding [24] can distort the spatial relationships, compromising the integrity of TME representation. Thus, integration absolute spatial context remains a key challenge for accurately measuring the spatial organization and interactions within TME to improve clinical prediction.

To address this challenge in computational pathology, we introduce EAGLE-Net, an **E**ffective **A**bsolute positional encoding and attention-**G**uided neighborhood-aware **L**oss **E**stimation **Network** designed to enhance foundation models' application. We demonstrate its performance through benchmarking on large pan-cancer datasets, including three cancer types for classification (totaling 10260 whole-slide images) and seven distinct cancer types for survival prediction (totaling 4,172 slides from 2,956 patients).

## METHODS

### EAGLE-Net Overview

EAGLE-Net is a MIL-based framework illustrated in **(Fig. 1)** that combines several key elements: (i) Tiling and feature extraction, in which tissue patches are extracted and embedded through the pretrained foundation model; (ii) Multi-scale Absolute Positional Encoding (MASE) block; (iii) Attention pooling; and (iv) Neighborhood-aware and background-suppression loss terms. We discuss these aspects in the following subsections. MASE module **(Fig. 1b)** aims to simultaneously learns patch-level information and global tissue structure using absolute positional encoding. Additionally, the proposed approach incorporates the neighborhood context of highly contributing patches in the training process to perform attention-driven profiling of relevant local regions **(Fig. 1d)**.

EAGLE-Net provides a distinct advantage over post-hoc analysis by learning clinically relevant tumor niches or clinically relevant regions during the model training using attention guidance. Furthermore, EAGLE-NET is compatible with any existing pre-trained foundation models. Unlike fixed coordinate systems, our method dynamically learns spatial context from histopathology data. Decoupling positional encoding from feature extraction can enhance spatial reasoning in existing models without modifying their architectures or pretrained weights.

**Pan-Cancer Datasets**

EAGLE-Net is comprehensively evaluated using seven prognostic and three diagnostic tasks of totally 14,432 WSIs from multiple institutions and scanners in 7 distinct cancer types. For survival analysis, we used six datasets sourced from The Cancer Genome Atlas (TCGA) and the Clinical Proteomic Tumor Analysis Consortium (CPTAC). From TCGA, we used lung squamous cell carcinoma (LUSC), lung adenocarcinoma (LUAD), stomach adenocarcinoma (STAD), Uterine Corpus Endometrial Carcinoma (UCEC), Thyroid Cancer Atlas (THCA), kidney renal clear cell carcinoma (KIRC) and CPTAC-LUAD (for details, see **Extended Data Fig. 1a-b**). These datasets cover a diverse set of cancer types. Experiments on these datasets are conducted using 5-fold Monte Carlo cross-validation, and average results are reported. We focused on predicting overall survival (OS), with the concordance index (C-index) as evaluation metrics.

For classification / subtyping, we performed experiments for both binary and multiclass classification. we used TCGA and CPTAC lung cancer subtyping (2 class). Similar to [16], we divided the data into train/validation/test sets with a ratio of 80%:10%:10% for both TCGA and CPTAC. We also evaluated the model's performance across different sets by locking models trained on TCGA and externally testing it on CPTAC data. Additionally, we performed multiclass classification of ISUP grades based on prostate cancer grade assessment (PANDA, 6-class) [25,26] (for details, see **Extended Data Fig. 1c**). PANDA comprises slides from Karolinska and Radboud Medical Centers. We performed training on combined data and evaluated on separate cohorts. We used balance accuracy as evaluation criteria for subtyping tasks, while Cohen's $\kappa$ for the ISUP grading task.

**Slide Processing and Patient-Level Tissue Packing**

Similar to [16], WSIs were patched at 20× magnification (0.5 μm/pixel), with a patch size of 256 × 256. Additionally, we cropped the tissue region from the slide and remove excessive background, minor artifacts, and empty regions between the tissues to reduce the size of the slide. For each patch, we extracted the feature using state-of-the-art UNI2-h foundation model [8], pre-trained on large-scale histology imaging datasets.

In patient-level analysis, a single patient often yields several distinct slides. In this case, instead of processing each slide in isolation, we performed a patient-level "tissue packing" step that packs tissue samples from multiple slides into one coherent canvas. For every slide, tissue patches were first extracted into a tightly cropped grid. These per-slide grids were then greedily rotated in quarter-turns to minimize their horizontal footprint and concatenated from left to right, forming a unified matrix whose rows and columns preserved the within-slide micro-architecture—allowing downstream visualization or slide-specific statistics when needed. The illustration of the packing approach is presented in (**Extended Data Fig. 1d**). For detailed tissue packing algorithms, see (**Extended Data Algorithm 1-2**). Packing patient-level slides onto a single grid gives EAGLE-Net the key advantage of a holistic view across multiple tissue samples of same patient, capturing inter-slide heterogeneity that would otherwise be neglected. Different tissue samples capturing a different facet of the tumor—an infiltrative front in one section, a lymphovascular nest in another—were analyzed together rather than in isolation. This unified canvas enabled the attention mechanism also to consider micro-lesions found on a specific tissue in conjunction with the dominant morphology of the primary resection, yielding a patient-level prediction that better reflects the full histological spectrum.

**Multi-scale Absolute Spatial Encoding (MASE)**

Inspired by the inherent modeling of the spatial locality and hierarchy by CNN [27], we proposed Multi-scale Absolute Spatial Encoding (MASE) module, which was designed to preserve global tissue structure by accounting for the position of adjacent tissues in the slide. Subsequently, MASE leveraged a two-stage convolutional approach to learn absolute positional encodings of tissue patches for every slide. The workflow of MASE is illustrated in (**Fig. 1b**).

Let $\mathcal{B}_i$ denotes the $i$-th WSI of size $h_i \times w_i$ in the dataset, we divide $\mathcal{B}_i$ into a set or bag of non-overlapping patches $\mathbf{X}_i = \{\mathbf{x}_{i,j}\}_{j=1}^{n_i}$, s.t. $\mathbf{x}_{i,j} \in \mathbb{R}^{h \times w}$, where ($w_i \gg w$ and $h_i \gg h$). Additionally, each bag $\mathbf{X}_i$ is associated with a bag-level label $\mathcal{Y}_i \in 0, \cdots, t$, where $t$ denotes the number of classes, and the label of individual patches remain unknown. To effectively capture the absolute spatial structure of the patches in $\mathbf{X}_i$, MASE uses set of patch location information in the WSI, $\mathbf{C}_i = \{(r_j, c_j)\}_{j=1}^{n_i}$, where $(r_j, c_j)$ is a unique tuple, s.t. $r_j \in \{1, r_i^*\}$ and $c_j \in \{1, c_i^*\}$, while $r_i^* = \left\lfloor \frac{w_i}{w} \right\rfloor$ and $c_i^* = \left\lfloor \frac{h_i}{h} \right\rfloor$ denote the total number of row-wise and column-wise patches in the slide, respectively. The patch location ($r_j, c_j$) contains the absolute position of patches $\mathbf{x}_{i,j}$ inside the WSI $\mathbf{X}_i$. We then employ a feature extractor $f_{\text{enc}}(:)$ pre-trained foundation model to extract patch features, $\mathbf{H}_i \leftarrow f_{enc}(\mathbf{X}_i)$, s.t $\mathbf{H}_i = \{h_{i,j}\}_{j=1}^{n_i}$ and $h_{i,j} \in \mathbb{R}^{1 \times m}$, where $m$ denotes the dimension of the feature. Additionally, we create an indicator vector $\mathbb{I}_i \in \mathbb{B}^{1 \times n_i}$ to segregate tissue and background patches as:

$$\underset{1 \leq j \leq n_i}{\forall} \mathbb{I}_i(j) = \begin{cases} 0, & \text{if } \mathbf{x}_{i,j} \text{ contains Tissue} \\ 1, & \text{otherwise} \end{cases}$$

Later, the indicator vector $\mathbb{I}_i$ is incorporated in the loss function for focused learning on tissue regions. Afterwards, the bag embedding $\mathbf{H}_i \in \mathbb{R}^{n_i \times m}$ are transformed to a position aware 3D representation matrix $\mathbf{M_i} \in \mathbb{R}^{r_i^* \times c_i^* \times d}$ as:

$$\underset{1 \leq j \leq n_i}{\forall} \underset{1 \leq k \leq m}{\forall} \mathbf{M}_i[(r_j, c_j), k] \leftarrow h_{i,j,k}$$

where $h_{i,j,k}$ is the element of $\mathbf{h}_{i,j}$ in $k$-th dimension. The transformation process ensures that the feature vectors in matrix $\mathbf{M}_i$ preserve the exact global tissue structure presented in slide $\mathcal{B}_i$. Later, we use $\mathbf{M}_i$ to learn absolute positional encodings.

In several sudies[24,27-29], it is demonstrated that convolution operation can effectively capture spatial positioning. Motivated by this fact, we proposed a two-stage convolution approach to effectively learn the absolute positional encodings for the bag. This method uses different-sized convolutional kernels to capture the bag's global spatial context at multiple scales.

In the first stage, we transform patch features to latent embeddings using a fully connected layer $f_e(:)$, s.t. $\mathbf{H}_i \leftarrow f_e(\mathbf{M}_i) \in \mathbb{R}^{r_i^* \times c_i^* \times d}$, where $(d < m)$. We then apply convolution layers with a kernel size of $1 \times 1, 3 \times 3, 5 \times 5$, and $7 \times 7$, using padding of $0, 1, 2$, and $3$, respectively. Small kernels encode fine spatial details, such as cellular structures, while larger kernels incorporate broader spatial context, like surrounding blood vessels and tissues, which help to capture heterogeneity in TME. The output feature maps of these convolution operations are stacked channel wise to capture diverse spatial structural patterns. In the second stage, we apply $1 \times 1$ convolution operation on the stacked feature map to retain absolute positional context while preserving spatial integrity. This two-stage convolution enhances the model's ability to capture complex spatial structure of tissues, which is expressed as:

$$\forall_{k \in \{1,3,5,7\}} \mathbf{V}_k = \text{Conv}(\mathbf{H}_i, \mathbf{W}_{k \times k}, b_k, \text{pad} = p),$$

$$\mathbf{V}_{\text{stack}} = \text{Concat}(\mathbf{V}_1, \mathbf{V}_3, \mathbf{V}_5, \mathbf{V}_7) \in \mathbb{R}^{r_i^* \times c_i^* \times 4 \cdot d},$$

$$\mathbf{P}_i = \text{Conv}(\mathbf{V}_{\text{stack}}, \mathbf{W}_{1 \times 1}, b_5, \text{pad} = 3) \in \mathbb{R}^{r_i^* \times c_i^* \times d},$$

$$\hat{\mathbf{H}}_i = \mathbf{H}_i \oplus \mathbf{P}_i,$$

where $k$ denotes the kernel size, and the padding $p$ are defined by $p = \frac{k-1}{2}$. We combine the latent representation of the bag $\mathbf{H}_i$ and obtain the absolute positional encoding matrix $\mathbf{P}_i$ to obtain a spatially enriched feature matrix $\hat{\mathbf{H}}_i$ for the bag. Finally, we reshape $\hat{\mathbf{H}}_i$ back to 2D representation to facilitate further MIL-based analysis:

$$\mathbf{Z}_i = \text{reshape}(\hat{\mathbf{H}}_i) \in \mathbb{R}^{n_i \times d}.$$

**Attention-based Neighborhood-aware Loss**

To directly integrate top-ranked tumor patches and their surrounding instances into the model's learning process, we proposed a novel neighborhood-aware loss term allowing the model to self-guide on clinically relevant niches. Specifically, we built upon the foundational work of loss-based attention for MIL [11,30], that links the cross-entropy based attention mechanism with the loss function by sharing weights of classification and attention layers, and proposed a novel neighborhood-aware loss that incorporates top attended instances and their connecting local neighborhood to bag-level loss function. The incorporation of local neighborhood awareness allows the model to extract variations in nearby regions rather than relying on a single patch. Biologically, local information around essential patches represents relevant niches that uncover subtle variations in TME and are critical for understanding tumor biology, such as growth, invasion, and therapeutic response.

Formally, let $f_\theta(:)$ be the classification layer, with $\mathbf{W} \in \mathbb{R}^{d \times t}$ weights, $b \in \mathbb{R}^t$ bias. By sharing the weights and bias of $f_\theta(:)$, the pooled bag representation vector $\mathbf{z}_i$ and attention weight of each instance $\mathbf{z}_{i,j} \in \mathbf{Z}_i$ can be computed as:

$$\forall_{1 \leq j \leq n_i} \alpha_{i,j} = \frac{\sum_{k=0}^{t-1} \exp(\mathbf{z}_{i,j} \mathbf{w}_k + b_k)}{\sum_{n=1}^{n_i} \sum_{q=0}^{t-1} \exp(\mathbf{z}_{i,n} \mathbf{w}_q + b_q)},$$

$$\mathbf{z}_i = \sum_{j=1}^{n_i} \alpha_{i,j} \mathbf{z}_{i,j},$$

where $\mathbf{w}_k$ denotes the $k$-th column vector from $\mathbf{W}$ and $b_k$ is the corresponding bias for $k$-th class. While $\alpha_{i,j}$ is the attention weight of instance $\mathbf{z}_{i,j}$ in the bag $\mathbf{Z}_i$.

In multiclass classification, the target vector is a one-hot encoding. Therefore, only the positive class contributes to the loss computation. Therefore, if bag $\mathbf{Z}_i$ belongs to the $k$-th class the equation for loss can be written as:

$$L_1 = -\log\left(\frac{\exp(\mathbf{z}_i\mathbf{w}_k + \mathbf{b}_k)}{\sum_{q=0}^{t-1}\exp(\mathbf{z}_i\mathbf{w}_q + \mathbf{b}_q)}\right).$$

$L_1$ denotes the task-specific bag-level loss, and the objective is to minimize this term, s.t. $L_1 \to 0$. However, attaining $L_1 \to 0$ does not guarantee that multiple instances are correctly weighted [48,60]. It is worth noting that instances with high attention scores are likely to serve as strong evidence supporting the bag's label, as attention scores are learned in a supervised manner using the slide-level labels during training. Therefore, highly attended instances and their neighborhood provide essential information regarding tumor behavior and can be used for attention-guided regional profiling.

Thus, we use the attention scores $\alpha_i$ to identify the top-k relevant instances and find their absolute local neighborhood in the generated 3D matrix representation $\hat{\mathbf{H}}_i$ in a given receptive field of radius $r$. Given the top-k instances with their neighborhood set and attention scores, the neighborhood-aware loss for bag $\mathbf{Z}_i$ is computed as:

$$L_2 = \sum_{v \in T}\sum_{j \in \mathcal{N}(\mathbf{z}_i, \mathbf{A}_i, v, r)} \left(-\log\left(\frac{\exp(\mathbf{z}_{i,j}\mathbf{w}_k + \mathbf{b}_k)}{\sum_{q=0}^{t-1}\exp(\mathbf{z}_{i,j}\mathbf{w}_q + \mathbf{b}_q)}\right)\alpha_{i,j}\right),$$

where $T$ denotes the set of the top-K attention wights for the instances in the bag $\mathbf{Z}_i$. While $\mathcal{N}(:)$ is a function that returns the corresponding neighborhood set of instances for the top weight $v \in T$ in matrix $\hat{\mathbf{H}}_i$. The detailed description of $\mathcal{N}$ is given in (**Algorithms 1–2**).

**Algorithm 1: Find Neighbors ($\mathbf{Z}_i$, $\hat{\mathbf{H}}_i$, $v, r$)**

    Input: $\hat{\mathbf{H}}_i$: $m \times n \times d$ spatially enriched feature tensor
    $\mathbf{Z}_i$ : bag of spatially enriched instances
    $v$ : linear index of one Top-K weight
    $r$ : receptive-field radius
    Output: $\gamma_i$: linear indices of neighbors in $\mathbf{Z}_i$

1   $I_v \leftarrow$ index ($v, \mathbf{Z}_i$)     *linear index of v in $Z_i$*
2   $m, n \leftarrow$ shape ($\hat{\mathbf{H}}_i$)
3   Row_Index $\leftarrow I_v$ div $n$
4   Col_Index $\leftarrow I_v$ mod $n$
5   (R, C) $\leftarrow$ Get Neighbors (Row_Index, Col_Index, $m, n, r$)
6   $\gamma_i \leftarrow \{ r_j \cdot n + c_j \mid (r_j, c_j) \in (R, C) \}$
7   return $\gamma_i$

**Algorithm 2: Get Neighbors (Row_Index, Col_Index, $m, n, r$)**

    Input: row Indices, column Indices: location in $\hat{H}_i$
    $m, n$: rows and columns of $\hat{H}_i$
    $r$: neighborhood radius

    Output: R: set of row indices; C: set of column indices

1   R $\leftarrow \emptyset$, C $\leftarrow \emptyset$
2   $minR \leftarrow \max(0, \text{Row\_Index} - r)$
3   $maxR \leftarrow \min(m - 1, \text{Row\_Index} + r)$
4   $minC \leftarrow \max(0, \text{Col\_Index} - r)$
5   $maxC \leftarrow \min(n - 1, \text{Col\_Index} + r)$
6   for i $\leftarrow$ minR to maxR do
7     for j $\leftarrow$ minC to maxC do
8       R $\leftarrow$ R $\cup \{i\}$;   C $\leftarrow$ C $\cup \{j\}$
9   return (R, C)

**Regularization to Mitigate Non-Tissue Background Effects**

Furthermore, while representing the bag as a 3D matrix, we may include some background or non-tissue patches to capture the absolute spatial structure. To eliminate the impact of non-tissue patches, we propose an additional regularization term:

$$L_3 = \sum_{j=1}^{n_i} \alpha_{i,j} \cdot \mathbb{I}_i(j).$$

$L_3$ term penalizes the attention weights of the background patches. This term helps the model weight relevant tissue patches and improve focus on tissue regions.

**Total Loss and Ablation Experiments**

The total loss to train the model is computed as:

$$\text{Loss} = L_1 + \lambda L_2 + \beta L_3,$$

where $\lambda$ and $\beta$ are user-defined hyperparameters. A large value of $\lambda$ presents a large penalty in terms of highly attended patches and their neighborhood. While a large value of $\beta$ encourages non-tissue instances to be driven toward zero. The loss computation steps for the proposed EAGLE-Net are illustrated in (**Fig. 1d)**.

Ablation experiments are planned as: (i) removing the MASE module; (ii) varying the top-k neighborhood size; (iii) adjusting the spatial radius of top ranked patches; and (iv) excluding L2 and L3 regularization losses. The detailed experiment designs are elaborated in Supplementary Methods (Ablation Experiments). We also provide a mathematical foundation related to the proposed loss function and a detailed ablation study for other hyperparameters is given in (**Extended Data Fig. 8)** and Supplementary material.

**Benchmark EAGLE-Net Against Existing MIL Algorithms on Pan-Cancer Tasks Across Multiple Backbone Foundation Models**

To evaluate performance, we benchmarked EAGLE-Net against several state-of-the-art supervised attention-based MIL methods, such as Attention-MIL (AbMIL)[14,30] and CLAM [15]. These methods assign attention weights to individual instances (patches) within a slide, generating slide-level representation via weighted averaging. Other methods, such as Attention-MISL [18] combines slide-level clustering with attention pooling. Additionally, Transformer-based MIL (TransMIL)[24] and low-rank MIL (ILRA) [17] introduce inter-instance correlation to enhance bag-level feature representations. Together, these benchmark methods represent leading MIL algorithms in computational pathology.

To assess the generalizability of EAGLE-Net beyond a specific foundation model, we conducted experiments across three widely used histology foundation models—REMEDIES [31], Uni-V1[8], and Uni2-h[8]. These foundation models provide diverse patch-level representations and differ in both training objectives and architectures. Our evaluation investigates whether EAGLE-Net's learned absolute spatial encodings and neighborhood-aware loss consistently provide performance gains across heterogeneous foundation backbones.

**Quantitative Evaluation of Interpretability and Biological Relevance**

We conducted a qualitative comparison of attention heatmaps to analyze different models' focus across the slides. Beyond qualitative and visual analyses, we carried the quantitative assessment by comparing attention maps with pathologist annotations. Specifically, we measured the distribution of attention across different tissue types to assess biological plausibility. A total of 300 TCGA-LUAD WSIs were annotated by in-house pathologists to generate high-resolution ground truth [32]. The annotation process and evaluation of biological relevance are illustrated in **Extended Data Fig. 2a-b**. To our knowledge, this is the first large scale, systematic assessment of attention map interpretability using comprehensive expert annotations. The annotations covered seven distinct biological regions—Tumor, Stroma, Immune, Vessel, Bronchi, Necrosis, and Lung [32], enabling direct comparison between model-derived attention and known histological structures.

We quantified the model's ability to identify the tumor region, using the Dice coefficient, false-positive rate, and frequency-domain descriptors—including Radial Energy Profile and Angular Energy Dispersion. To complement spatial-domain evaluation, we introduced two frequency-domain descriptors derived from the squared magnitude of the 2-D Fourier transform $|F(u,v)|^2$ of the tumor mask, where $u$ and $v$ are spatial-frequency coordinates:

1. Angular Energy Dispersion (AED):

    AED characterizes the anisotropy in frequency space by partitioning the spectrum into $N$ angular set of sectors $\Psi = \{E_1, E_2, E_3, \ldots, E_N\}$ where energy in each sector $E_i$ is computed as:

    $$E_i = \sum_{(u,v) \in \text{sector} i} |F(u,v)|^2,$$

    and AED is defined as the Shannon entropy over the angular energy distribution:

$$p_i = \frac{E_i}{\sum_{i=1}^{N} E_i},$$

$$\text{AED} = -\sum_{i=1}^{N} p_i \log p_i$$

Higher AED values indicate dispersed directional energy, capturing spiculated and irregularly contour margins typically associated with invasive and heterogeneous tumors. Conversely, low AED reflects isotropic, smoothly contoured shapes.

2. Radial Energy Profile:

Radial Energy Profile assesses how spectral energy decays with frequency. The Fourier spectrum is divided into set of $M$ concentric annuli $\Phi = \{R_1, R_2, R_3, \ldots, R_M\}$, and the energy in each annulus $R_j$ is defined as:

$$R_j = \sum_{(u,v) \in \text{annulus } j} |F(u,v)|^2,$$

$$\tilde{R}_j = \frac{R_j}{\sum_{j=1}^{M} R_j}.$$

This profile reveals the distribution of spatial detail: a sharply peaked $\tilde{R}_j$ centered at low frequencies signifies smooth, well-circumscribed masks, while broader distributions indicate high-frequency components corresponding to fine-grained irregularities and complex contours.

Both AED and the radial profile are invariant to translation, rotation, and isotropic scaling, and inherently suppress pixel-level noise. Together, they provide a compact, biologically meaningful representation of tumor shape. Visual example of radial and angular energy profile is shown in **Extended Data Fig. 2c**.

# RESULTS

**Innovations of EAGLE-Net Compared to Existing Attention-Based MIL Models**

As illustrated in **Extended Data Fig. 3**, we benchmark EAGLE-Net against seven MIL frameworks—TransMIL[24], ILRA[17], Att-MISL[18], AbMIL[14], Gated-AbMIL[14], CLAM[15], DSMIL[33]—using following four essential standards:

*Instance-level Explainability*: All compared MIL models offer instance-level explainability through attention mechanisms, enabling the visualization of diagnostically relevant regions in whole-slide images, which is crucial capability for clinical integration. For instance, TransMIL[24] employs self-attention to evaluate patch relevance; CLAM applies clustering-constrained attention for instance weighting; AbMIL and Gated-AbMIL[14] provides attention-based pooling; DSMIL features a dual-stream architecture for critical instance mining; and Att-MISL implements attention-guided selection learning. EAGLE-Net enhances interpretability by sharing weights between the classification and attention branches, thereby aligning model prediction with visual saliency. However, several methods such as ILRA, Att-MISL, AbMIL, and CLAM operate without incorporating any spatial context of the tissue structure.

*Attention-guided region profiling:* EAGLE-Net uniquely supports attention-guided neighborhood profiling, enabling it to infer context around high-attention patches in a self-supervised manner. This is particularly valuable in histopathology, where biologically relevant features (e.g., immune infiltration, tumor-stroma boundaries) span beyond individual patches. By incorporating Multi-scale Adaptive Spatial Encoding (MASE) and a neighborhood-aware loss, EAGLE-Net captures inter-regional dependencies that are critical for accurate characterization of the tumor microenvironment—a capability absent in other MIL models.

*Positional encoding:* Among the evaluated models, only EAGLE-Net and TransMIL explicitly model spatial information. TransMIL approximates positional information by reshaping the bag into 2D grids for transformer-based encoding. However, this reshaping may distort the true neighborhood structure of the patches in the original slide. In contrast, EAGLE-Net introduces dual-scale spatial encoding in MASE (**Fig. 1b**). This design allows the model to learn spatial patterns reflective of histological structures, enhancing its ability to identify contextual biomarkers and structural features of the TME.

*Bag-Wise Instance Correlations:* Both TransMIL and ILRA model inter-instance dependencies within a bag—TransMIL through self-attention across the slide, and ILRA[17] via instance-level graph. However, these methods fall short in capturing fine-grained spatial organization. EAGLE-Net addresses this limitation by leveraging the hierarchical structure and spatial inductive bias of CNNs, enabling it to model multi-scale spatio-temporal correlation with spatial inductive bias, which reflects cross-regional tissue microarchitecture. This hierarchical modeling facilitates learning of multi-scale spatial patterns critical for interpreting tumor heterogeneity, tissue interfaces, and cellular gradients in a biologically meaningful manner.

**EAGLE-Net Achieves Superior Performance on Pan-Cancer Prognostic Tasks**

To evaluate the prognostic utility of EAGLE-Net's slide-level representations, we applied Cox proportional-hazards models to attention-pooled embeddings and computed concordance indices (C-index) using 5-fold Monte Carlo cross-validation. EAGLE-Net was benchmarked against widely used attention-based MIL architectures—including AbMIL, Gated-AbMIL, CLAM, TransMIL, ILRA, Att-MISL, DSMIL across six TCGA cohorts and CPTAC-LUAD cohort (4,172 slides from 2,956 patients, see **Extended Data Fig. 1a-b**). EAGLE-Net consistently achieved improved or comparable prognostic performance compared to benchmarked algorithms as shown in **Fig. 2a**.

In TCGA-KIRC, EAGLE-Net achieved C-index of $0.708 \pm 0.018$, surpassing CLAM ($0.668 \pm 0.026$), Gated-AbMIL ($0.693 \pm 0.008$), AbMIL ($0.687 \pm 0.010$), while matching TransMIL ($0.705 \pm 0.005$). For TCGA-LUAD, EAGLE-Net scored $0.672 \pm 0.018$, ahead of CLAM ($0.662 \pm 0.012$), TransMIL ($0.650 \pm 0.016$), Gated-AbMIL ($0.657 \pm 0.020$), and AbMIL ($0.631 \pm 0.025$). In TCGA-LUSC, EAGLE-Net attained $0.690 \pm 0.018$, outperforming CLAM ($0.683 \pm 0.005$), TransMIL ($0.683 \pm 0.017$), Gated-AbMIL ($0.685 \pm 0.006$), and AbMIL ($0.681 \pm 0.008$). On TCGA-STAD, EAGLE-Net achieved $0.697 \pm 0.010$, outperforming CLAM ($0.687 \pm 0.024$) and TransMIL ($0.664 \pm 0.031$). For TCGA-THCA, EAGLE-Net scored $0.688 \pm 0.005$, exceeding CLAM ($0.670 \pm 0.011$) and AbMIL ($0.679 \pm 0.011$). In TCGA-UCEC, EAGLE-Net reached $0.706 \pm 0.031$, comparable to CLAM ($0.707 \pm 0.009$) and TransMIL ($0.712 \pm 0.025$), but higher than Gated-AbMIL ($0.689 \pm 0.047$) and AbMIL ($0.694 \pm 0.041$). On CPTAC-LUAD cohort, EAGLE-Net achieved $0.723 \pm 0.052$, outperforming all other models. Overall, EAGLE-Net demonstrated consistently strong prognostic capability across diverse tumor types with average c-indices ranging from 0.685 to 0.743 (**Fig. 2a)**.

**EAGLE-Net Reaches Robust Classification Performance**

EAGLE-Net was benchmarked on two NSCLC cohorts for histology classification: the TCGA dataset (n = 1043; LUAD vs. LUSC) and the CPTAC cohort (n = 785) as external validation. EAGLE-Net achieved Accuracy of 0.980 on TCGA and 0.916 on CPTAC **(Fig. 2b)**. In comparison, CLAM reached 0.980 and 0.889, ILRA achieved 0.927 and 0.862, and TransMIL yielded 0.972 and 0.892, respectively.

To assess generalizability, we extended EAGLE-Net to multiclass classification of ISUP prostate cancer grades using the PANDA dataset, consisting of 6,548 training slides, 895 validation slides, and two held-out test cohorts from Karolinska (n = 481) and Radboud (n = 418). On the Radboud cohort, EAGLE-Net archived a Cohan's kappa of 0.984, outperforming CLAM 0.970 and TransMIL (0.944, **Fig. 2c**). On the Karolinska cohort, EAGLE-Net achieved a Cohan's kappa of 0.985, comparable to CLAM (0.985) and exceeding TransMIL (0.954, **Fig. 2c**). These results underscore EAGLE-Net's ability to learn robust positional encoding and morphological representations that generalize across variations in staining intensity and scanner hardware from multicenter data.

**EAGLE-Net Provides Cross-Backbone Consistency Across Different Foundation Models**

To demonstrate the generalizability of EAGLE-Net beyond specific foundation models, we conducted experiments using three popular histology foundation models—REMEDIES[31], Uni-V1[34], and Uni2-h[34]—which differ in architectural design and training objectives. These models generate diverse patch-level embeddings, allowing us to assess whether EAGLE-Net's spatial encoding and neighborhood-aware loss functions retain their benefits regardless of the underlying feature extractor.

For each backbone, we trained EAGLE-Net, CLAM, Gated-AbMIL, and AbMIL from scratch on four TCGA cohorts (LUAD, KIRC, LUSC, and STAD). EAGLE-Net consistently achieved the highest C-index across all three backbones and cancer types (**Fig. 3**). Using REMEDIES[31] (**Fig. 3a-b**), EAGLE-Net's surpassed the next-best model by 0.1–3.9%, exceeded the cohort mean by 1.8–6.0%, and outperformed the weakest competitor by 3.2–7.6%. With Uni-V1 (**Fig. 3c-d**),

EAGLE-Net's performance gains ranged from 0.2–1.7% over the runner-up, 0.6–2.6% over the mean, and 1.2–3.4% above the lowest-performing baseline. Similarly, under Uni2-h (Fig. 3e-f), EAGLE-Net maintained its lead with 0.3–1.1% over the next-best model, 0.9–2.3% above the average, and 1.2–4.2% over the lowest baseline. These results highlight EAGLE-Net's robustness and adaptability, demonstrating consistent performance advantages across foundation models trained with different objectives—including contrastive learning (REMEDIES) and self-distillation with augmentation invariance (Uni-V1, Uni2-h).

**Attention Heatmaps Provide Insights into EAGLE-Net's Decision-Making**

To evaluate the interpretability of EAGLE-Net, we analyzed attention heatmaps and compared their spatial correspondence with hallmark histomorphological features (Fig. 4). At the global level, EAGLE-Net's attention map exhibits smoother, more continuous saliency contours that align well with tumor boundaries and stromal interfaces—suggesting a coherent, context-aware representation of tissue architecture. What distinguish EAGLE-Net from other attention-based models is its ability to assign consistent attention scores to spatially contiguous regions exhibiting similar microstructural patterns. This enhances both biological plausibility and interpretability, as local regions with analogous morphology are weighted similarly. Such behavior supports improved predictive performance and contributes to model trustworthiness by reflecting tumor heterogeneity in a biologically consistent manner. Notably, EAGLE-Net effectively identifies clinically relevant tumor subregions without requiring explicit region-of-interest (ROI) annotations. The resulting attention maps highlight its potential to uncover biologically and clinically significant niches that might be missed by the fragmented and inconsistent attention patterns of other models.

Extended comparisons across four datasets (TCGA-LUAD, TCGA-LUSC, CPTAC-LUAD and CPTAC-LUSC) are shown in Extended Data Figs. 4-7. In each case, EAGLE-Net outperforms Gated-AbMIL and CLAM in aligning saliency maps with coarse expert annotations, consistently emphasizing tumor-associated structures by assigning nearly identical attention scores to morphologically similar areas.

**Inter-model Attention Correlation Analysis Reveals Distinctive Feature Patterns**

To further characterize the uniqueness of EAGLE-Net's attention mechanism, we designed an inter-model correlation framework to compare its patch-level attention scores with those from CLAM, Gated-AbMIL, and AbMIL. Pairwise similarity was measured using the mean Pearson correlation of attention scores across all patches as well as the intersection-over-union ($IoU_{10}$) of the top 10% most highly attended patches. This dual-metric approach enables comparison of both continuous attention distributions and discrete high-saliency regions **(Fig. 5a-b)**.

The analysis revealed that EAGLE-Net shares moderate correlation with Gated-AbMIL (mean r ≈ 0.63) and lower correlation with CLAM (r ≈ 0.38, **Fig. 5c-d**), highlighting architectural differences in how spatial attention is assigned—particularly to epithelial regions. The modest inter-model correlations, coupled with EAGLE-Net's superior performance, suggest that beyond key histomorphological patterns revealed by prior MIL, it reveals new spatial patterns. This may contribute to improved boundary delineation and enhanced prognostic accuracy.

**EAGLE-Net's Generated Attention Maps Align with Tumor Masks**

We assessed the biological relevance of EAGLE-Net's attention maps by thresholding them into predicted tumor masks and comparing these with expert-annotated tumor boundaries at the patch level using the Dice similarity coefficient, false-positive rate (FPR), and frequency-domain metrics **(Fig. 6a)**. This evaluation provided a comprehensive measure of the model's ability to accurately localize diagnostically relevant tissue regions. EAGLE-Net achieved the highest Dice score (0.56), suggesting superior alignment between predicted tumor regions with ground truth compared to other attention-based methods. It also achieved the lowest FPR (0.101), reflecting fewer false highlights of non-tumor regions. In frequency-domain analysis, EAGLE-Net achieved a lower radial difference of 0.147 and a lower AED Difference of 0.316, demonstrating robustness in delineating tumor boundaries and capturing patterns of invasiveness.

We performed statistical comparisons among EAGLE-Net, CLAM, and Gated-AbMIL using four matrices—Dice coefficient, FPR, radial difference, and AED, with two-sided Wilcoxon signed-rank tests with Bonferroni correction ($\alpha = 4.2 \times 10^{-3}$) and reporting effect sizes as median differences. EAGLE-Net showed statistically significant improvement in Dice score over CLAM and in FDR over both CLAM and Gated-AbMIL (**Fig. 6b**). For frequency domain metrics, EAGLE-Net is not statistically significant in radial difference from CLAM; however, it exhibits statistically significant improvements over Gated-AbMIL, with similar patterns observed for AED.

**Overlap of EAGLE-Net's Attention with Fine-Grained Manual Annotations**

We further examined the distribution of attention across pathologist-annotated tissue regions by calculating the proportion of total attention allocated to each region within a given WSI. Attention allocation patterns varied across architectures, reflecting different computational strategies for sampling diagnostic cues within the heterogeneous LUAD microenvironment (Fig. 7a-g). EAGLE-Net allocates 74.6% of its attention to tumor regions, exceeding CLAM (73.5%) and Gated-AbMIL (72.8%), thereby sharpening sensitivity to neoplastic tissue. Necrotic areas received 2.40% attention (vs. 2.25% for CLAM and 2.33% for Gated-AbMIL), capturing the necrotic heterogeneity often associated with aggressive disease. Immune infiltration was also better highlighted, with lymphocytes receive 0.893% of attention compared with 0.774% (CLAM) and 0.863% (Gated-AbMIL). In contrast, attention to normal lung regions was reduced (6.10% vs. 8.20% for CLAM and 7.84% for Gated-AbMIL), as was attention to vascular structures (2.93% vs. 3.39% and 3.33%). A modest increase in attention to bronchial recognition (0.558% vs. 0.515% and 0.542%) preserves the key architectural context. Overall, these shifts in attention indicate that EAGLE-Net more effectively targets diagnostically relevant compartments—tumor, necrosis, and immune infiltrates—while de-emphasizing benign tissue, thereby improving focus on biologically meaningful regions and potentially facilitating biomarker discovery from histopathology slides.

**DISCUSSION**

We developed EAGLE-Net, an end-to-end trainable multiple instance learning (MIL) framework that integrates Multi-scale Absolute Spatial Encoding, a top-k neighborhood-aware loss, and background suppression loss to jointly model global tissue architecture and local microenvironment context from whole-slide images. By capturing absolute spatial positioning and emphasizing cohesive, biologically relevant niches, EAGLE-Net overcomes key limitations of existing weakly supervised computational pathology models. Across diverse TCGA and CPTAC cohorts and multiple histology foundation models, EAGLE-Net consistently outperformed state-of-the-art attention-based algorithms in classification and survival prediction, generating smooth, coherent attention maps that aligned with expert annotations, improved tumor boundary delineation, and reduced false positives. Dual-domain evaluation—combining spatial and frequency-domain metrics—further confirmed that EAGLE-Net preserves biologically plausible morphologies patterns.

From an engineering perspective, EAGLE-Net is foundation model–agnostic, enabling seamless integration with evolving histology backbones without re-designing the entire pipeline. This adaptability positions it for long-term sustainability as digital pathology datasets and pre-trained models continue to expand. It advances computational pathology in three key ways. First, it learns spatially aware attention without the need for dense annotations, making it practical for deployment in large-scale, resource-limited settings. Second, the top-k neighborhood-aware loss focuses attention on cohesive tumor microenvironments, rather than isolated high-scoring tiles, enhancing both interpretability and prognostic relevance. Third, its heatmaps provide histology-level explanations that align with clinical reasoning, enabling pathologists to visualize the model's decision-making process and supporting explainable AI objectives that are increasingly important for regulatory adoption.

By bridging the gap between weak supervision and biologically grounded interpretability, EAGLE-Net provides a generalizable platform for precision oncology. Future directions include multi-modal fusion of histology, imaging, and molecular biomarkers; real-time integration into digital pathology workflows; and prospective validation in clinical trials. These steps will be critical for advancing EAGLE-Net from a research framework to a clinically deployed system that augments pathologist expertise and improves patient outcomes.

Limitations remain. The Multi-scale Absolute Spatial Encoding module increases computational overhead during training, which could be mitigated through lighter spatial encoding schemes or hierarchical architectures. While EAGLE-Net excelled in large surgical specimens, performance gains were less pronounced in biopsy datasets such as PANDA, where spatial context is limited and morphological patterns overlap (e.g., Gleason grading). Frequency-domain metrics emphasize global shape but may overlook cell-level cues, suggesting the need for multi-resolution extensions to capture fine-grained cellular signals.

In summary, by uniting global spatial modeling with fine-grained microenvironment profiling, EAGLE-Net produces biologically faithful attention maps and robust predictions across cancer types. Its adaptability, interpretability, and strong performance across various foundation models position it as a versatile tool for biomarker discovery, patient stratification, and the development of next-generation AI systems in translational oncology.

**Code availability**

All code was implemented in Python using PyTorch as the primary deep-learning library. The complete pipeline for processing WSIs as well as training and evaluating the deep-learning models will be available at: https://github.com/WuLabMDA/EAGLE-Net

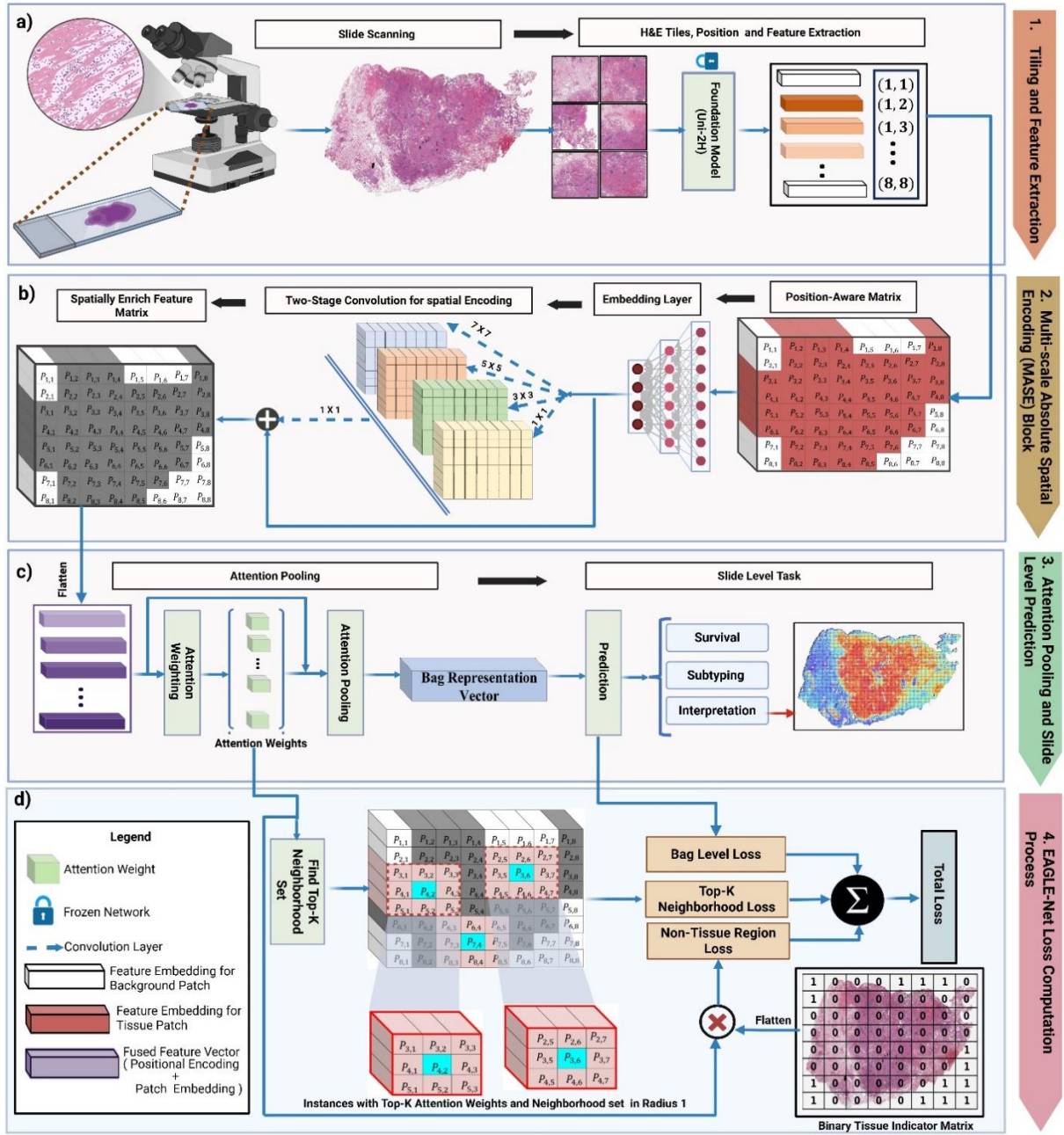

**Fig. 1: Overview of the proposed methodology**. **a,** preprocessing pipeline, including tissue segmentation, patch and feature extraction from whole-slide images. **b,** Architecture of the proposed MASE module employing two-stage convolutional approach to learn positional encodings and contextual information. **c,** the application of loss-based attention pooling for survival prediction, subtyping, or generation of interpretable attention maps highlighting diagnostically relevant regions. **d,** Training strategy illustrating the integrated proposed loss

function, combining slide-level classification loss with Top-K instance neighborhood loss and non-tissue region loss for improved biological alignment.

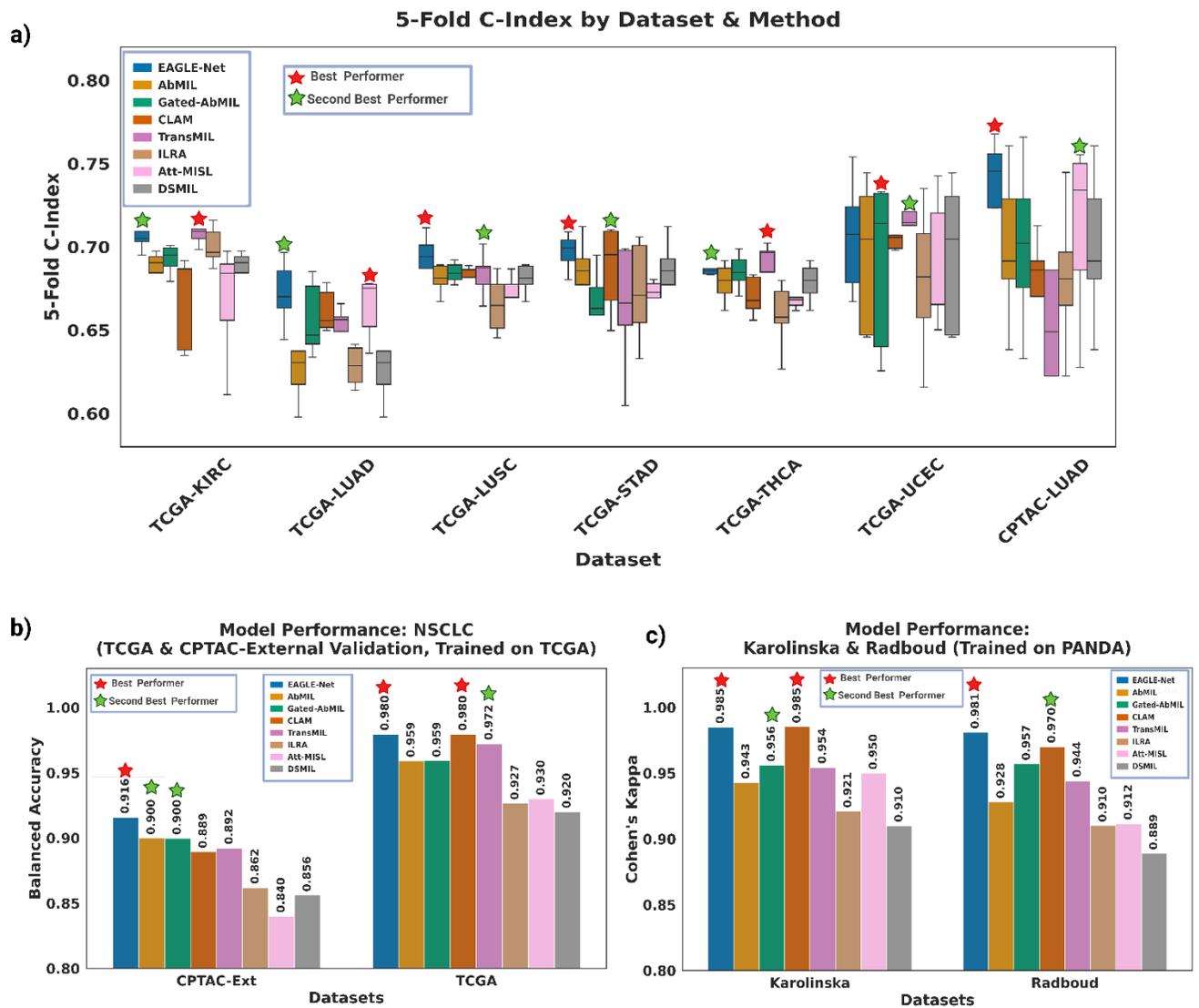

**Fig. 2: Experimental results for prognosis and diagnostic tasks.**

**a,** Survival analysis results showing concordance index (c-index) for 5-fold Monte-Carolo cross-validation across models for different cancer subtypes. **b,c,** Slide-level classification performance across multiple benchmark datasets.

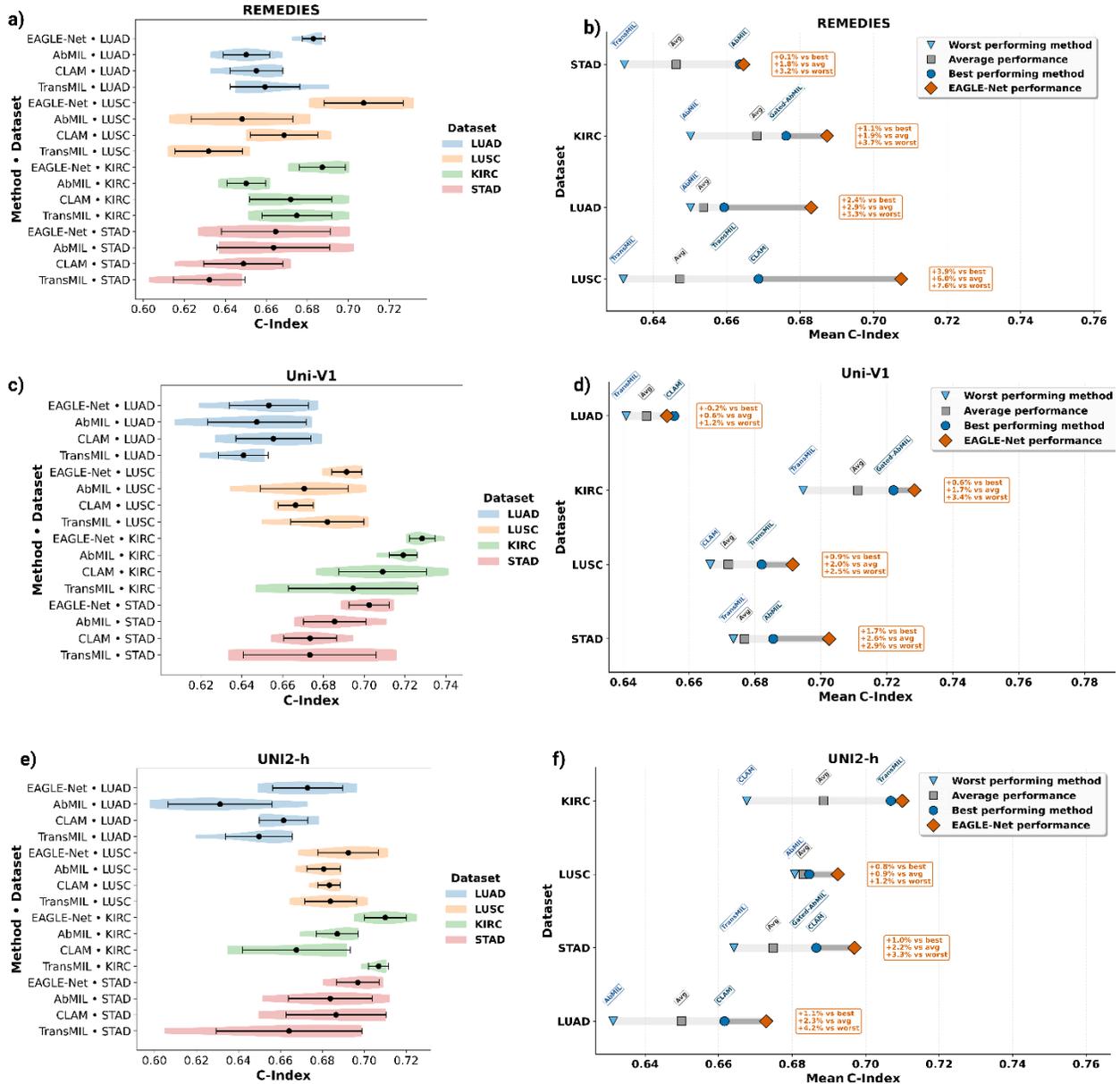

**Fig. 3: Encoder-Agnostic Prognostic Performance Across Self-Supervised Histology Backbones.**

Violin plots of ten-fold cross-validated concordance indices (C-index) for EAGLE-Net (Proposed) and three baseline weakly supervised models (CLAM, Gated-AbMIL and AbMIL) on four TCGA cohorts (LUAD, LUSC, KIRC and STAD), using three distinct self-supervised encoders: **a, b** REMEDIES, **c, d** Uni-V1 and **e, f** Uni2-h. Each violin reflects the full distribution of C-indices across folds, with white circles marking the median and black bars indicating the

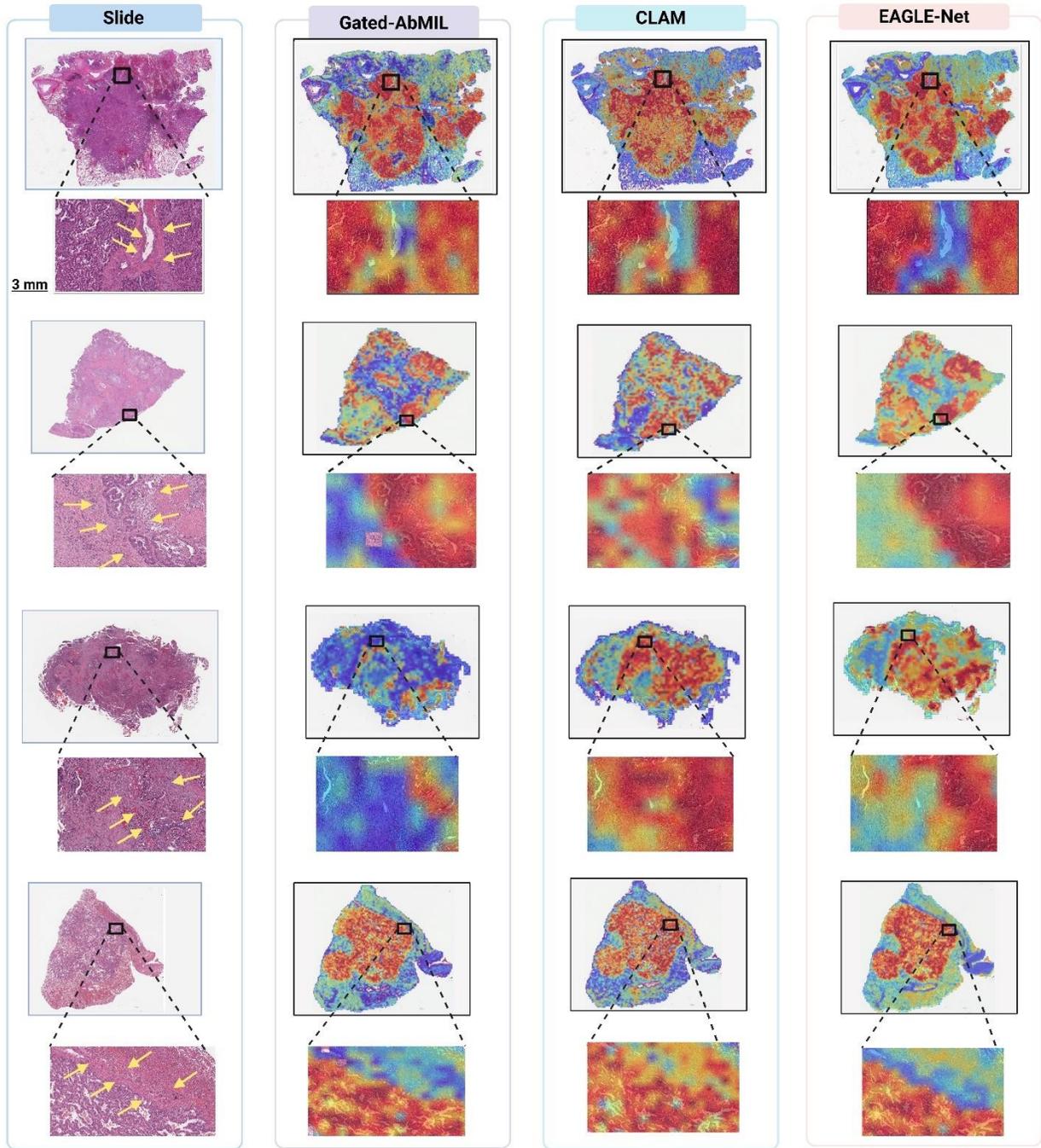

**Fig. 4: Comparative Attention Maps Across MIL Models with Zoomed-in Insets**

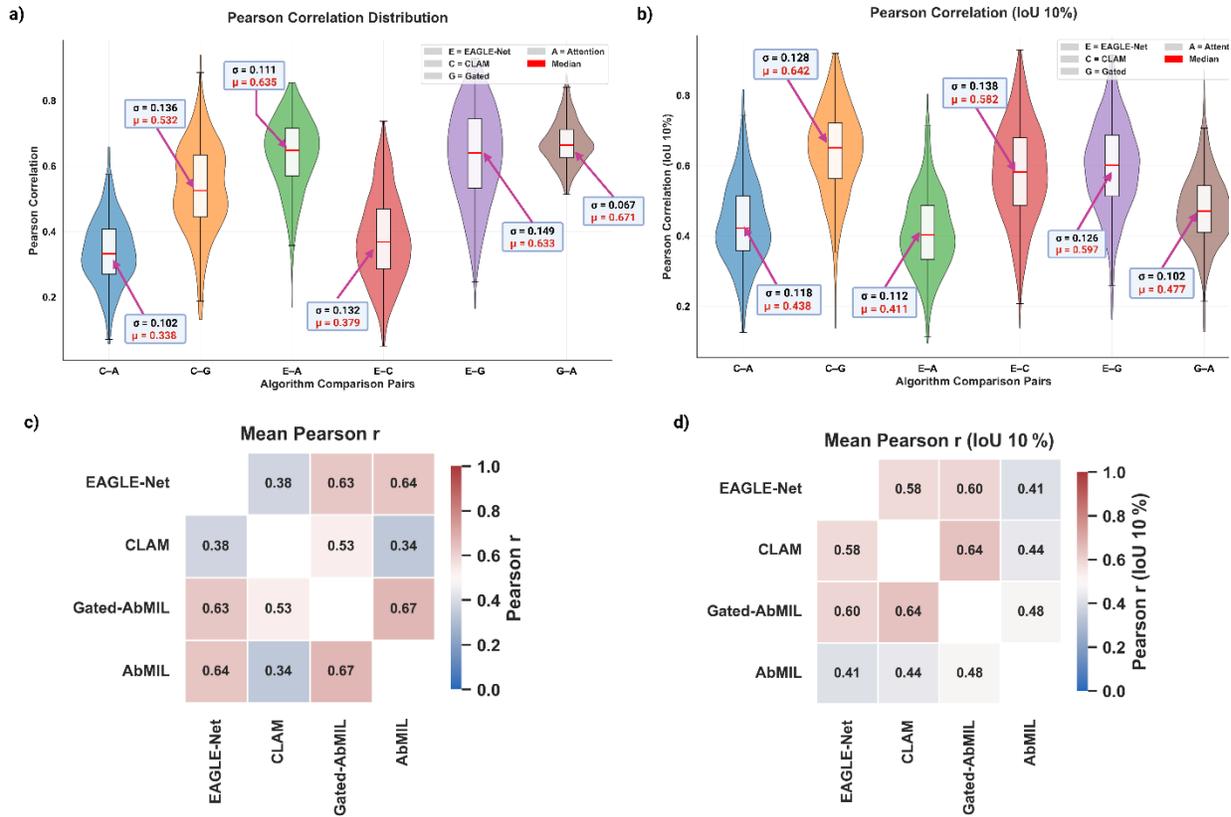

**Fig. 5: Inter-model attention concordance frameworks. a**, Violin plots of patch-level Pearson attention score correlation coefficients and correlation among scores for the top 10 % most-attended patches (intersection-over-union (IoU$_{10}$)), computed pairwise among EAGLE-Net, CLAM, Gated-AbMIL, and AbMIL. Boxes denote median ± interquartile range. **b**, Cohort-averaged heatmaps of mean Pearson r (left) and mean IoU$_{10}$ (right) for the same model pairs.

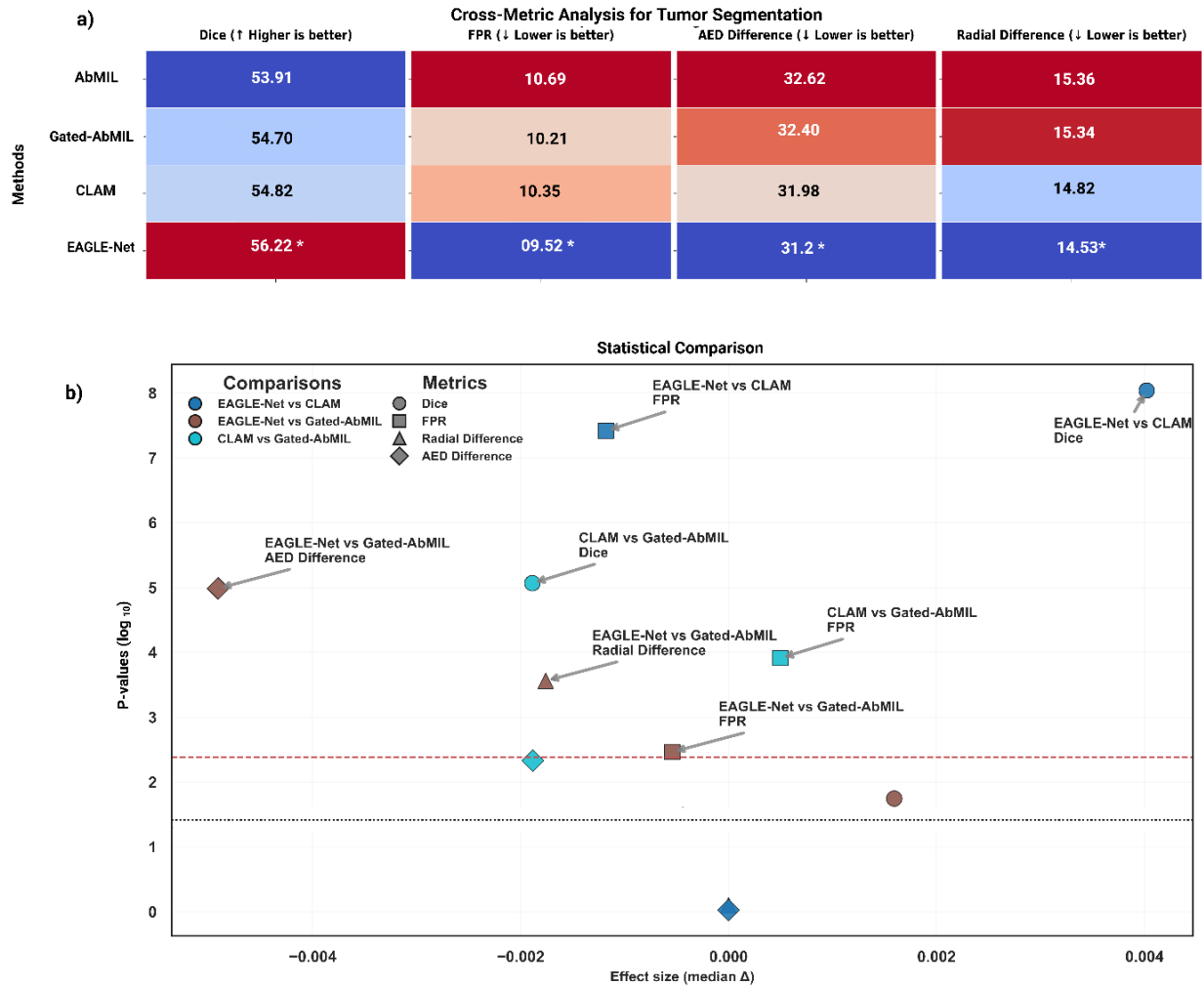

**Fig. 6: Qualitative performance, statistical analysis and attention mass allocation.**

**a**, Heatmaps of four boundary-segmentation metrics—Dice, FPR, AED and Radial difference, averaged over the cohort for each method. **b**, Slide-paired Wilcoxon signed-rank tests. Each point encodes one model–metric comparison: the x-coordinate is the median effect size (Δ = Proposed – baseline), and the y-coordinate is –$\log_{10}$(p-value). Marker shapes denote metrics and colors denote method pairs. The dashed red line marks the Bonferroni-corrected significance threshold (α = 0.05/12; –$\log_{10}$ ≈ 2.32), and the grey dotted line the nominal p = 0.05 cutoff (–$\log_{10}$ ≈ 1.30). Points above the red line are significant after correction and are labelled in the plot.

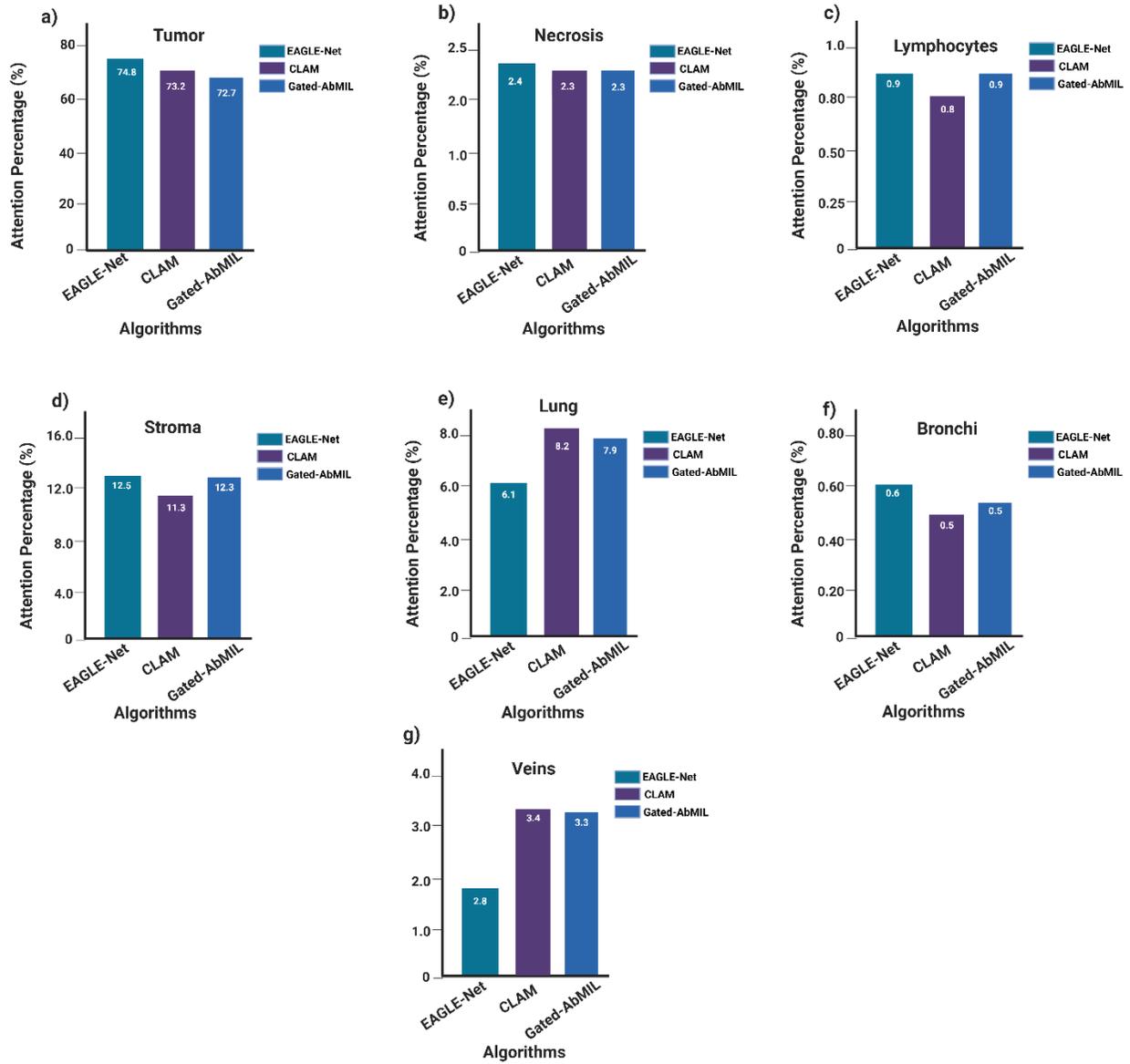

**Fig. 7: Comparison of attention allocation to different biological compartments.**

Average attention allocation to different tissue compartments (**a.** Tumor, **b.** Necrotic, **c.** Immune, **d.** Stromal, **e.** Lung, **f.** Bronchial, and **g.** Veins).

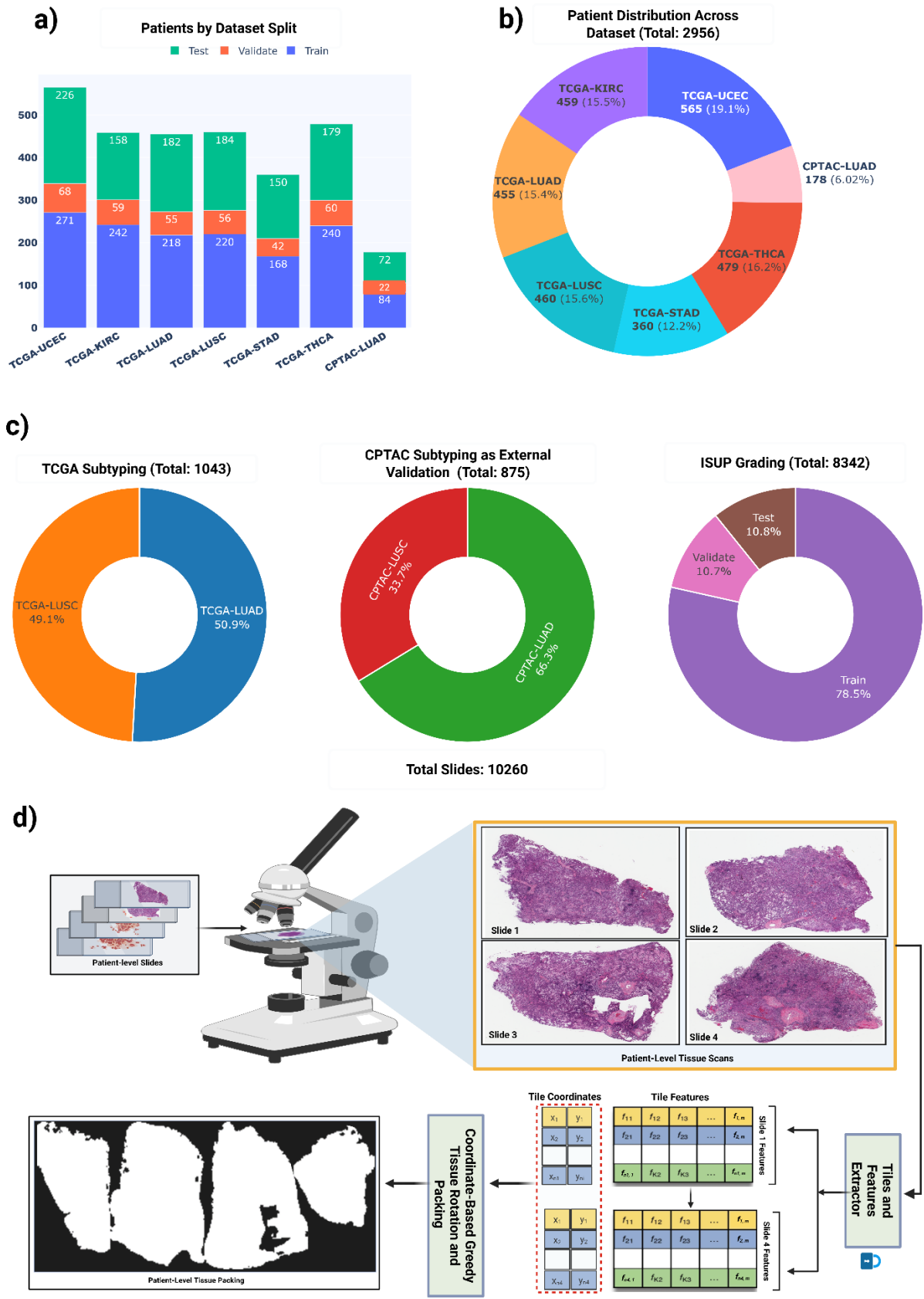

**Extended Data Figure 1: Overview of datasets and tissue-packing strategy. a, b** Pan-cancer cohort assembly for prognostic modelling, comprising six TCGA tumour types (KIRC, LUAD, LUSC, STAD, THCA, UCEC) and the independent CPTAC-LUAD dataset, with slide counts and patient metadata indicated for each. **c**, Summary of subtyping task datasets, and class distribution within datasets used for slide-level classification tasks. **c**, Classification and subtyping tasks: slide counts and class distributions for clear-cell renal-cell carcinoma subtyping, non-small-cell lung-cancer subtyping and lymph-node-metastasis detection. **d**, patient-level tissue packing on-the-fly from the tile coordinates output by the foundation model, embedding packing into the dataloader rather than a separate preprocessing step and keeping it plug-and-play with any model.

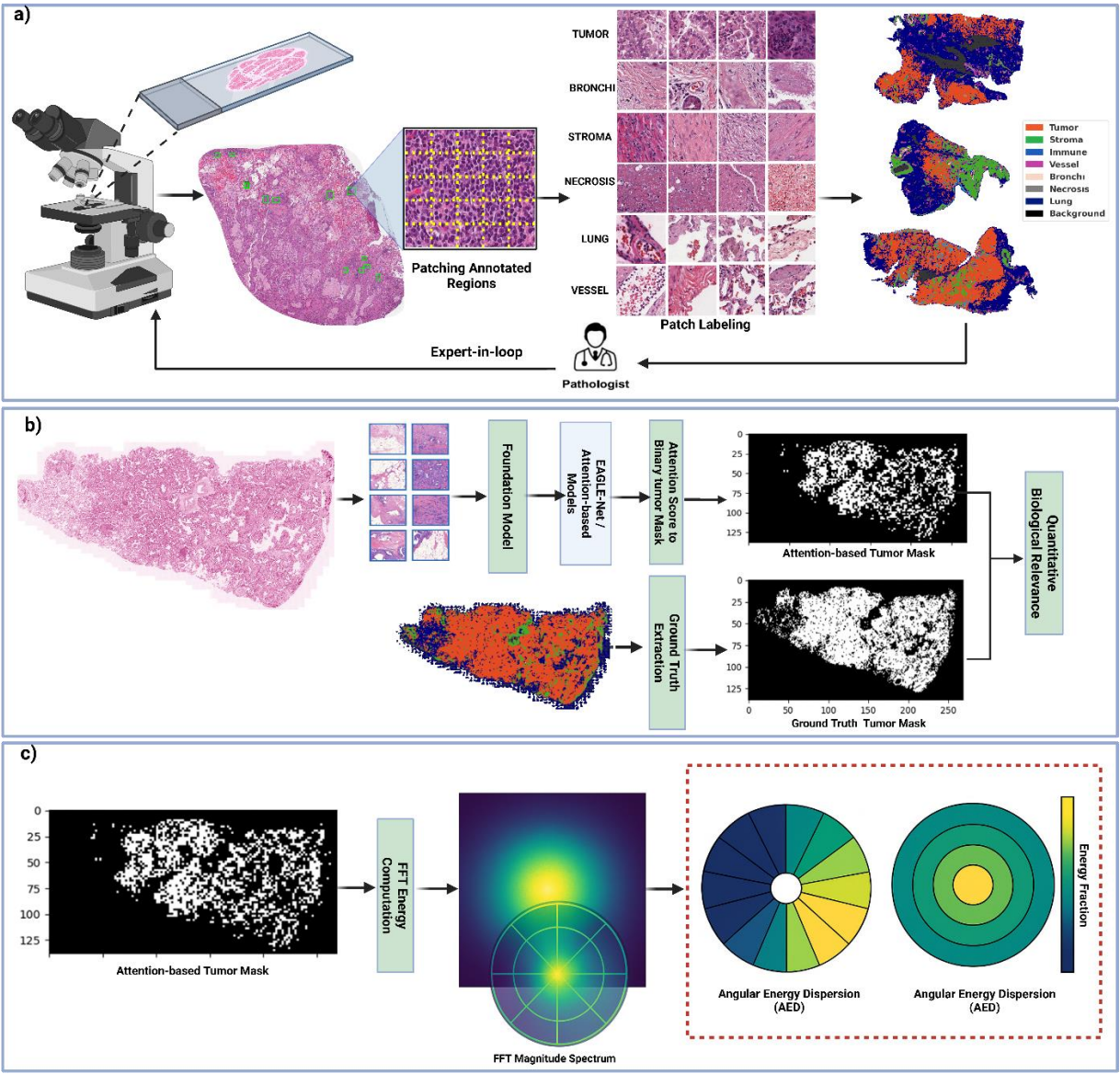

**Extended Data Figure 2: Patch-level annotation and evaluation pipeline. a,** Expert-guided annotation process for labeling tumor and non-tumor regions at the patch level across whole-slide images, used to generate high-resolution spatial ground truth. **b,** Transformation of model-derived attention maps into binary tumor masks, followed by quantitative comparison with expert annotations to assess biological relevance and localization accuracy. **c,** Schematic illustration of spectral decomposition into angular sectors and radial annuli for frequency-domain analysis.

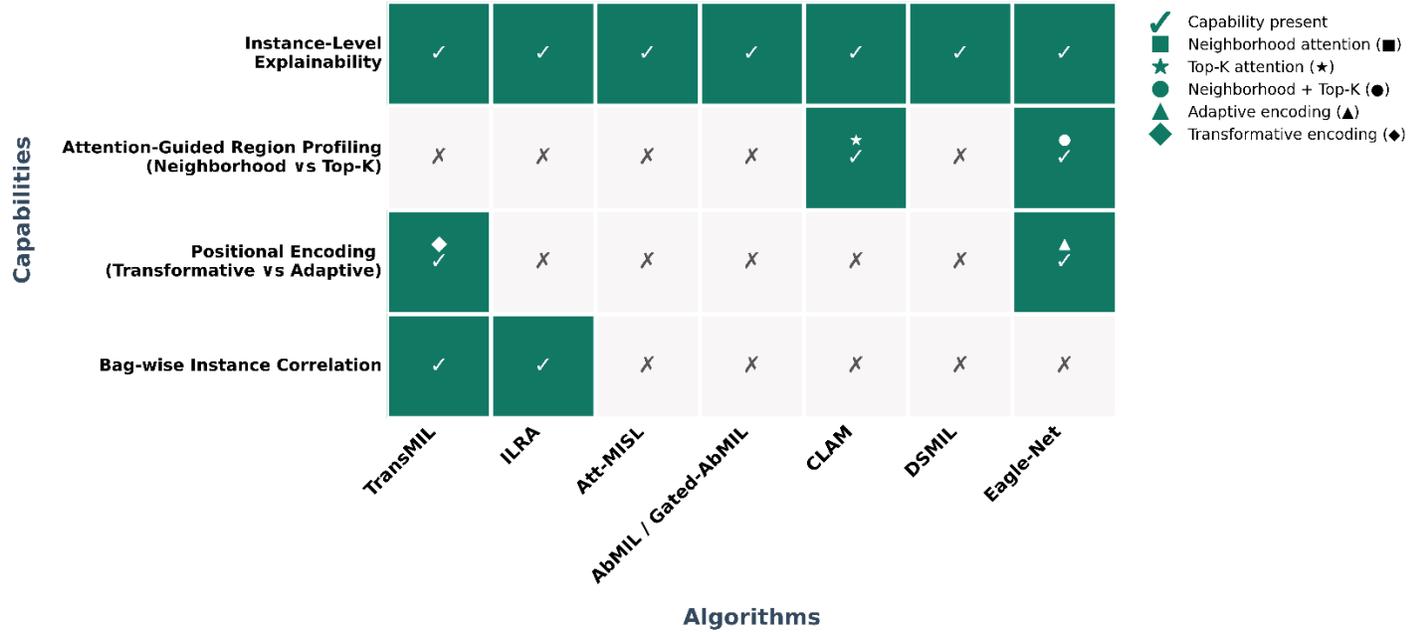

**Extended Data Figure 3: Comparative analysis of weakly supervised (MIL) algorithms for WSI analysis.**

Each row corresponds to a key architectural capability of MIL frameworks. A ✓ marks support for a given feature, while a ✗ denotes its absence. EAGLE-Net is the only method to unify most capabilities, illustrating its unique ability to generate compact attention maps alongside robust slide-level prediction.

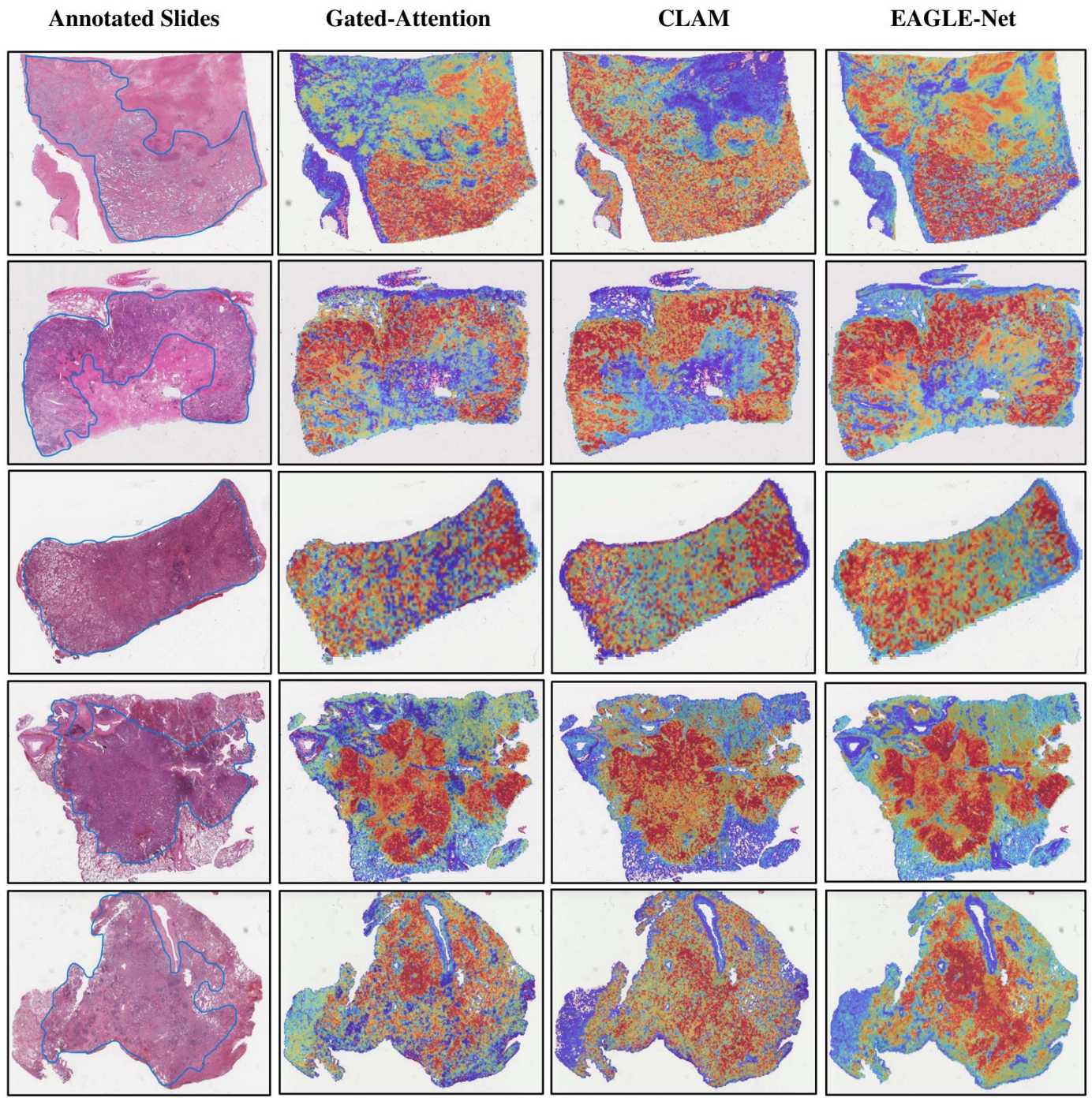

**Extended Data Figure 4: Comparative attention heatmaps for TCGA-LUAD slides.**

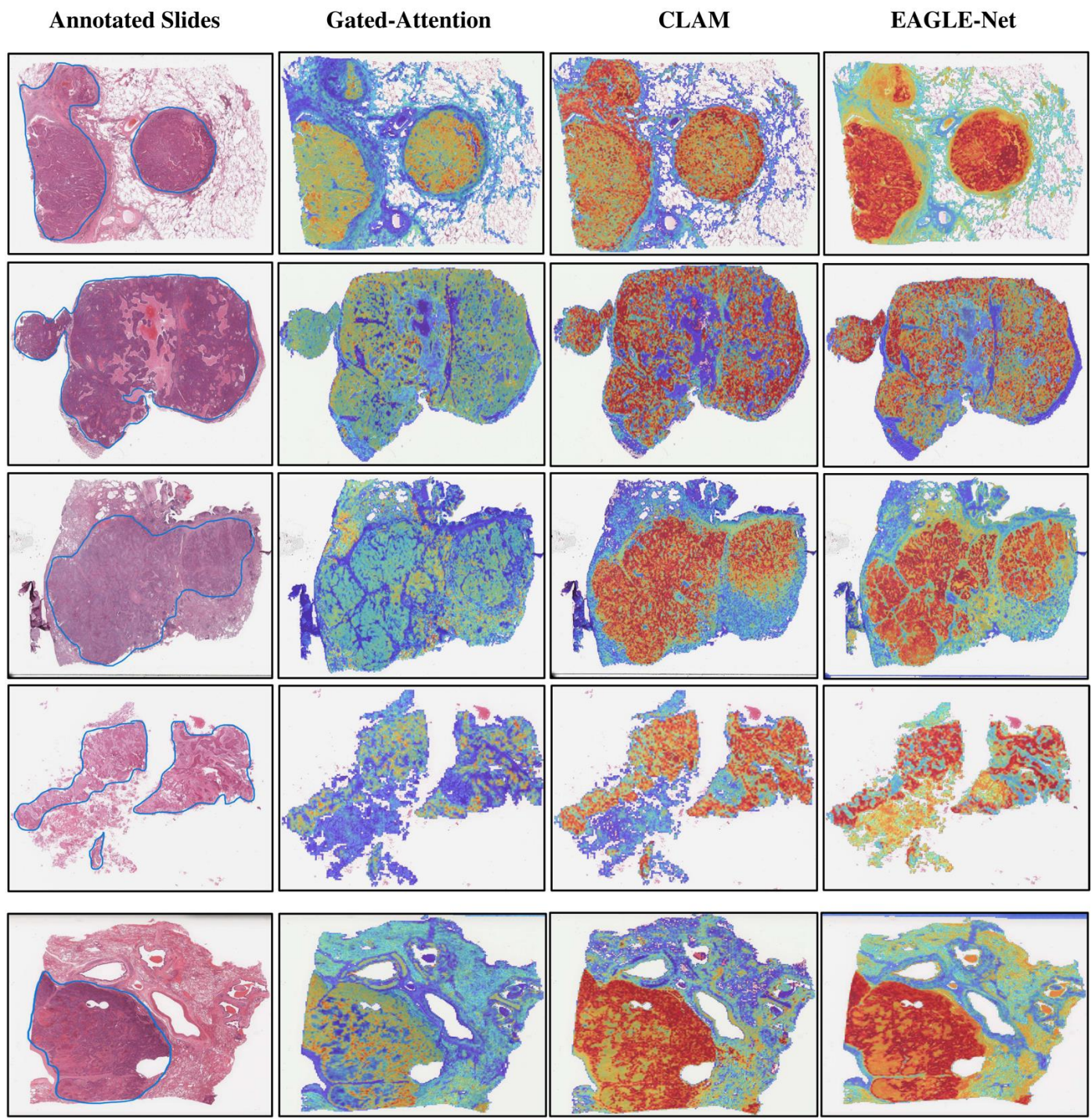

**Extended Data Figure 5:** Comparative attention heatmaps for TCGA-LUSC slides.

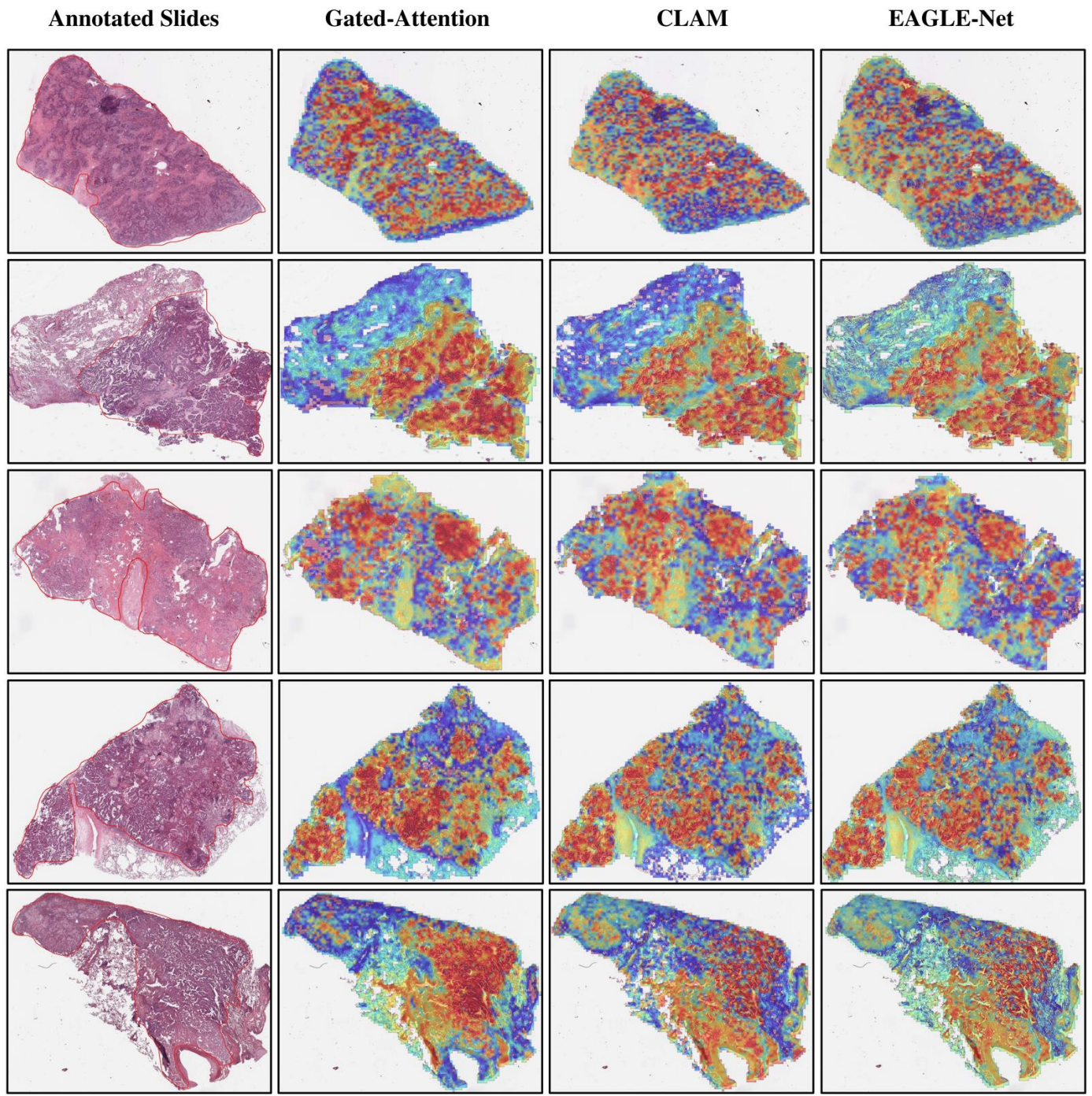

**Extended Data Figure 6: Comparative attention heatmaps for CPTAC-LUAD slides.**

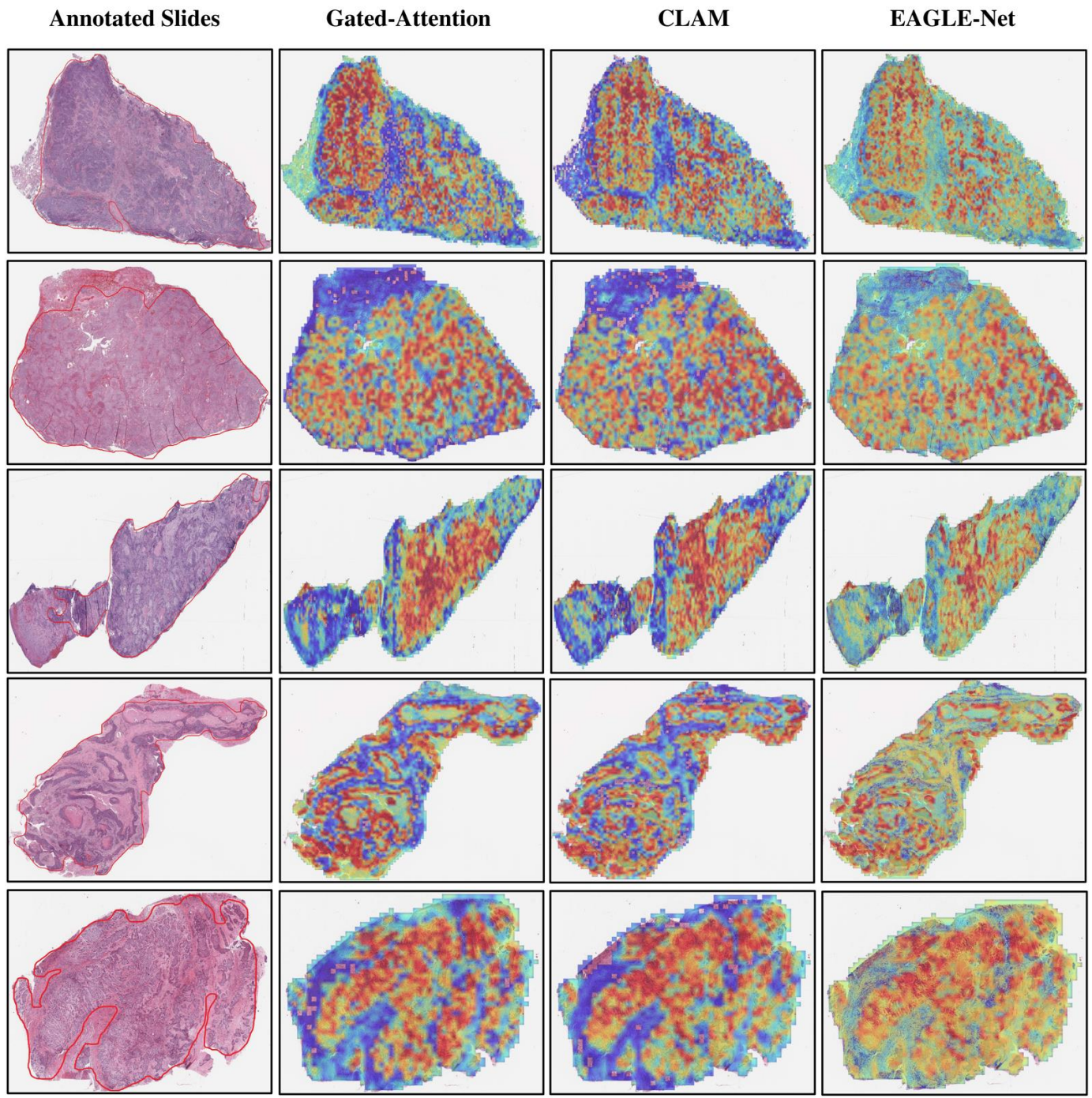

**Extended Data Figure 7: Comparative attention heatmaps for CPTAC-LUSC slides.**

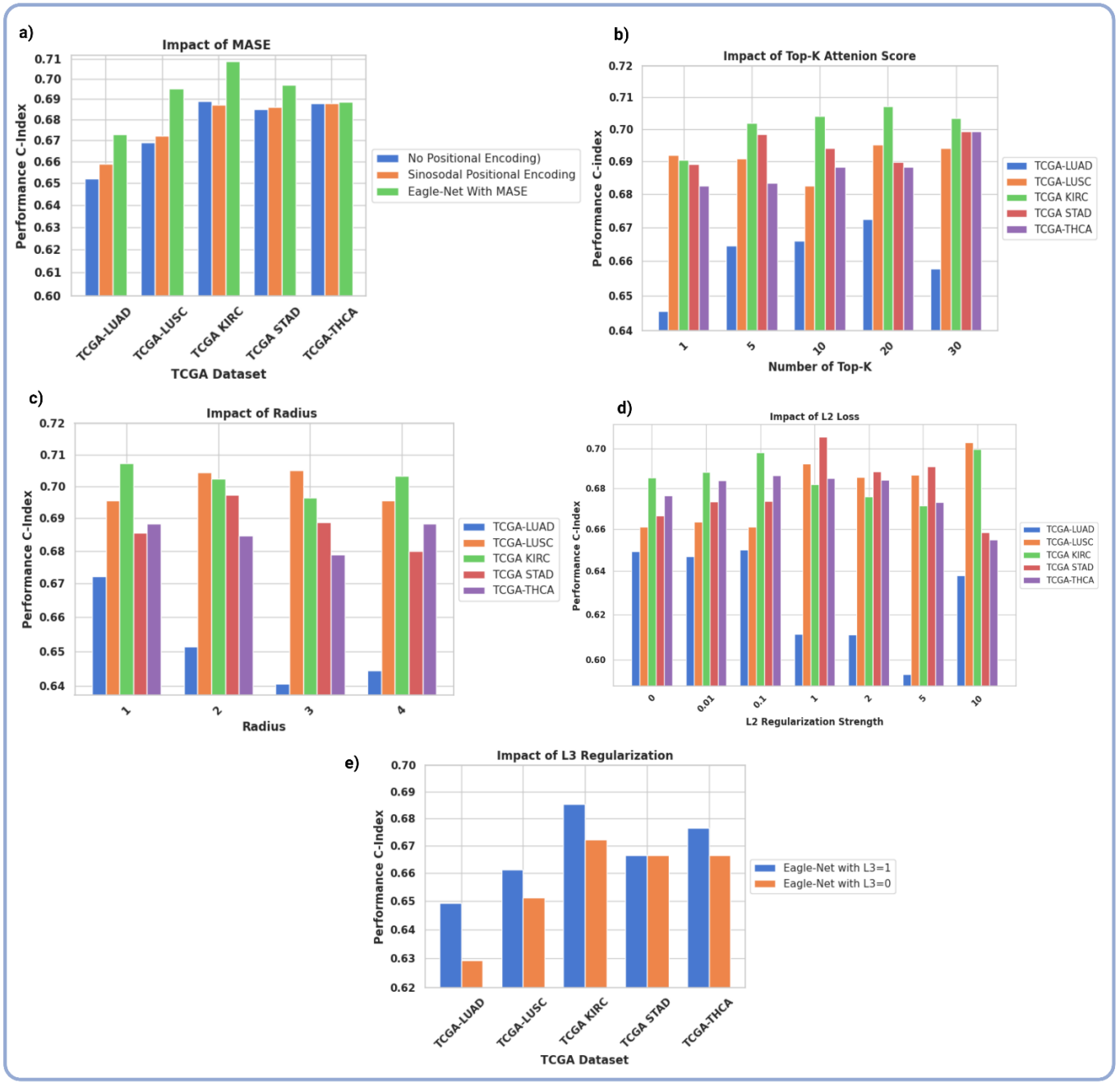

**Extended Data Figure 8**: **Ablation study of model components on survival prediction (c-index) across five TCGA cohorts.**

Survival concordance is shown for: **(a)** positional encoding scheme (none, sinusoidal, MASE); **(b)** number of top-attended patches (K); **(c)** local neighborhood radius (r); **(d)** auxiliary neighborhood consistency loss; and **(e)** background suppression loss.

**Extended Data Algorithm 1: PROCESSPATIENT(files, slideThr = 0.01, patientThr = 0.25)**

**Input**: *files: list of HDF5 paths; slide Thr: per-slide filtering threshold (default 0.01); patientThr: final patient grid threshold (default 0.25)*

**Output**: *PMmat: patient position-aware matrix; PMmask: patient mask*

1 **Load & preprocess each slide**
   slides ← [];
   foreach fp ∈ files do
     open fp; slides.append(PROCESSSLIDE(features, coords, slideThr));

2 **Greedy rotation to minimise horizontal area**
   slideMats, slideMasks ← [], [];
   W, H ← 0, 0;
   foreach (SM, MASK) ∈ slides do
     best ← argmin_{k∈{0,1,2,3}} area after k quarter-rotations;
     (SM*, MASK*) ← best;
     slideMats.append(SM*);
     slideMasks.append(MASK*);
     W ← W + SM*[1];   H ← max(H, SM*[0]);

3 **Horizontal concatenation & labelling**
   D ← feature_dim(slideMats[0]);
   PMmat ← $0^{\{H \times W \times D\}}$,  PMmask ← $0^{\{H \times W\}}$;
   col ← 0;
   for idx, (SM, MASK) ∈ enumerate(slideMats, slideMasks) do
     h,w ← MASK.shape;
     PMmat[:h, col:col+w, :] ← SM;
     PMmask[:h, col:col+w] ← (MASK > 0) × (idx + 1);
     col ← col + w;

4 **Final content-based cropping**
   rowThr ← patientThr × W;   colThr ← patientThr × H;
   rows ← { i | sum(PMmask[i, :]) ≥ rowThr };
   cols ← { j | sum(PMmask[:, j]) ≥ colThr };
   return PMmat[rows][:, cols], PMmask[rows][:, cols];

**Extended Data Algorithm 1**: **Patient-Level Tissue Packing Algorithm.**

Pseudocode for the generation of patient-level slide organization into a unified "packed" representation for EAGLE-Net. The algorithm enumerates and loads each WSI, rotate them to different angles to ensure that packing area is minimized and to merge multi-slide data into a single contiguous matrix that preserves inter-slide spatial context for downstream task.

**Extended Data Algorithm 2: PROCESSSLIDE (feats, coords, thr = 0.1, scale = 256)**

*Input*: feats: 1×N×D feature array; coords: N×2 raw patch coordinates;
thr: row/col activity threshold (default 0.1);
scale: coordinate scaling factor (default 256)

*Output*: SM: slide position-aware matrix; MASK: corresponding binary mask

1. **Shift & scale coordinates**
   x, y ← coords[:,0], coords[:,1]
   $x_0, y_0$ ← min(x), min(y);
   pts ← [ (x − $x_0$), (y − $y_0$) ] / scale ;
   H, W ← max(pts[:,1]) + 1, max(pts[:,0]) + 1;

2. **Initial mask**
   MASK ← $0^{\{H \times W\}}$;
   for each p ∈ pts do
   　　MASK[$p_y, p_x$] ← 1;

3. **Remove low-information rows / cols**
   rowThr ← thr × W;　colThr ← thr × H;
   rows ← { i | sum(MASK[i, :]) ≥ rowThr };
   cols ← { j | sum(MASK[:, j]) ≥ colThr };
   rm ← dict(enumerate(rows));　cm ← dict(enumerate(cols));

4. **Remap coordinates into filtered grid**
   new ← [ cm[$p_x$], rm[$p_y$] ] for each p;

5. **Populate position-aware matrix**
   f ← feats[0];
   D ← feature_dim(f);　// drop batch dim
   SM ← $0^{\{|rows| \times |cols| \times D\}}$,　MASK ← $0^{\{|rows| \times |cols|\}}$;
   foreach ($n_y, n_x$) ∈ new do
   　　SM[$n_y, n_x$, :] ← f;　MASK[$n_y, n_x$] ← 1;
   return SM, MASK;

**Extended Data Algorithm 2**: **Patient-Level Tissue Packing Algorithm.**

Pseudocode for the processing each slide, and generation of its tissue mask.